\documentclass[12pt]{article}
\usepackage{amssymb}
\usepackage{mathbbold}
\usepackage{amsfonts}
\usepackage{amsmath}
\usepackage{natbib,graphicx,setspace,lscape,longtable}
\usepackage{natbib,epsfig,graphicx}
\usepackage{mathrsfs,amsmath,amsthm,amssymb,color}
\usepackage{bm}
\usepackage{extarrows}
\usepackage{multirow}
\usepackage{cases}
 \usepackage{ctable}
\usepackage{booktabs}
\usepackage{threeparttable}
\usepackage{setspace}
\usepackage{algorithm}
\usepackage{algorithmic}
\usepackage{amssymb}
\usepackage{pdflscape}
\usepackage{subfigure}

\allowdisplaybreaks[4]
\makeatletter
\newcommand{\distas}[1]{\mathbin{\overset{#1}{\kern\z@\sim}}}%
\newsavebox{\mybox}\newsavebox{\mysim}
\newcommand{\distras}[1]{%
  \savebox{\mybox}{\hbox{\kern3pt$\scriptstyle#1$\kern3pt}}%
  \savebox{\mysim}{\hbox{$\sim$}}%
  \mathbin{\overset{#1}{\kern\z@\resizebox{\wd\mybox}{\ht\mysim}{$\sim$}}}%
}
\makeatother
\bibpunct{(}{)}{;}{a}{,}{,}

\newtheorem{theorem}{Theorem}
\newtheorem{lemma}{Lemma}
\newtheorem{proposition}{Proposition}

\newcommand{\csection}[1]
    {\begin{center}
        \stepcounter{section}
        {\bf\large\arabic{section}. #1}
    \end{center}
    \vspace{-0.15 cm}
}

\newcommand{\scsection}[1]
    {\begin{center}
        {\bf\large #1}
    \end{center}
    \vspace{-0.15 cm}
}

\newcommand{\csubsection}[1]{
\vspace{-0.25 cm}
\begin{center}
\stepcounter{subsection}
{\it\arabic{section}.\arabic{subsection}. #1}
\end{center}
\vspace{-0.25 cm}
}

\def\beq{\begin{equation}}
\def\eeq{\end{equation}}
\def\beqr{\begin{eqnarray}}
\def\eeqr{\end{eqnarray}}
\def\beqrs{\begin{eqnarray*}}
\def\eeqrs{\end{eqnarray*}}
\def\bet{\begin{theorem}}
\def\eet{\end{theorem}}
\def\bel{\begin{lemma}}
\def\eel{\end{lemma}}
\def\bep{\begin{proposition}}
\def\eep{\end{proposition}}
\def\bg{\begin{figure}[tbph]\begin{center}}
\def\eg{\end{center}\end{figure}}

\def\bc{\begin{center}}
\def\ec{\end{center}}

\def\diag{\mbox{diag}}

\def\tr{{\rm tr}}
\def\vec{{\rm vec}}

\numberwithin{equation}{section}

\begin{document}

\begin{center}
\Large \textbf{A two-way factor model for high-dimensional matrix data}
\end{center}
\vspace{1cm}
\begin{singlespace}
\vspace{-1cm}
%\begin{center}
$Zhigen\ Gao^{{1,{\dagger}}}$, $Chaofeng\ Yuan^{{1,2,{\dagger}}}$, $Bingyi \ Jing^{3}$, $Wei\ Huang^{1,*}$, $Jianhua\ Guo^{1,*}$
\footnotetext[2]{These authors contributed equally to the work}
\footnotetext[1]{Please address correspondence to huangw482@nenu.edu.cn and jhguo@nenu.edu.cn}
{\it $^1$KLAS and School of Mathematics and Statistics, Northeast Normal University.\\
     $^2$School of Mathematical Science, Heilongjiang University.\\
     $^3$Department of Mathematics, Hong Kong University of Science and Technology.}

%\end{center}
\begin{abstract}
In this article, we introduce a two-way factor model for a
high-dimensional data matrix and study the properties of
the maximum likelihood estimation (MLE). The proposed model
assumes separable effects of row and column attributes and captures
the correlation across rows and columns with low-dimensional
hidden factors. The model inherits the dimension-reduction
feature of classical factor models but introduces a new
framework with separable row and column factors, representing
the covariance or correlation structure in the data matrix.
We propose a block alternating, maximizing strategy to compute
the MLE of factor loadings as well as other model parameters.
We discuss model identifiability, obtain consistency and the
asymptotic distribution for the MLE as the numbers of rows and
columns in the data matrix increase. One interesting phenomenon
that we learned from our analysis is that the variance of the estimates
in the two-way factor model depends on the distance of variances
of row factors and column factors in a way that is not expected
in classical factor analysis. We further demonstrate the performance
of the proposed method through simulation and real data analysis.
\end{abstract}
\end{singlespace}

\newpage

\begin{singlespace}

\csection{Introduction}

The factor model, as a classical model in multivariate statistics, has been widely used
in the undertaking of high-dimensional data analysis in a variety of scientific
areas including finance, psychology, biology.
By introducing latent variables (known as factors), a
factor model assumes that observed variables are
independent from each other after being introduced to small numbers of latent factors and, as a result,
provides a simplified but meaningful framework to summarize
the effects of latent factors as well as covariance structures between observed variables.

Conventional factor-model-based methods
focus mainly on analyzing vector-valued data, in which the observable attributes
are converted into a vector, and the relevant observations
are considered to be independent or approximately independent samples
(\cite{Anderson:(1956)}; \cite{Anderson:(1988)}, et al.).
In recent studies,
\cite{Bai:2003}, \cite{Bai:2012}, \cite{Fan:(2008)} and \cite{Fan:(2011)}
have further generalized these methods under a high-dimensional framework and
obtained a series of remarkable theoretical results.
Nevertheless, these methods are faced with difficulties when applied to
matrix data analysis. This is mainly because
(a) there are often no replicates for
an observation of matrix-based variables;
(b) relationships between the attributes of rows and between those of columns should be considered separately;
(c) vectorization procedure in matrix-based observations usually
ignores distinguishing information about row and column, and
thus cannot separate the effects of attributes along rows and columns.
In terms of statistical methods that have been developed to address these issues,
\cite{Gupta:2000},
\cite{Werner:2008}, \cite{Leng:2012}
and \cite{Ding:2018}
focused mainly on a matrix data set with replicates.
\cite{Tsai:(2016)} considered doubly constrained factor models for a data matrix.
Though common factors could be interpreted by measuring effect of row,
column and interaction according to their models,
the doubly factor models indeed deal with vector-valued data
if the known constrained matrices are absorbed by common factors or factor loadings.
\cite{Wang Dong:2016}, \cite{Wang:2019} and \cite{Chen:2019}
proposed factor-model-based methods
for high-dimensional matrix-valued data with replicates,
in which a low-sized matrix is used to represent
the structures of hidden factors. However, their methods
do not make an explicit separation between the row and the column effect
and therefore can only provide evaluations of their joint behaviors.
Faced with the reality of no replicates of matrix data,
\cite{Zhou:2014} and \cite{Michael:2018} proposed two
methods which are derived
from a general matrix variate distribution, but they additionally
require sparsity assumptions about the covariance matrix in order to obtain the parameter estimations
as well as the large sample properties. Moreover, their ideas are essentially
not factor domain.

In this work, we extend the idea of factor models to the analysis of
high-dimensional matrix data, such as $X_{p \times q}$,
with no replicates. To the best of our knowledge, this is the first attempt to apply factor analysis to studying complex
correlation structures of matrix-wise variables with a single observation.
Our interest stems from an environmental study in which volume readings
of 14 chemicals (columns), including $SO_2$, $CO$, collected from
338 cities (rows), are reported and the aim is to discover any
patterns exhibited by cities and pollutants that
could provide a systematic explanation of the status of the air,
especially when extreme air pollution occurs.
This motivates us to consider a comprehensive method that can simultaneously
model possible pollution patterns through latent factors,
decompose the information from the data matrix by rows and columns, and
evaluate the effect of row and column factors based on the observed data matrix.

The core thinking behind our model is that the behaviors of each entry or variable in the data matrix $X$ are
affected by two groups of latent factors. One group
summarizes the effects of the row attributes, while the other summarizes the effects of column attributes.
The complex correlated relationships among the entries can then simply be decomposed and explained by the latent row and
column factors. More specifically,
we can assume that the data matrix $X$ can be regarded as a sum of two unobservable matrices
$$X = U + V,$$
where $U$ is the `row effect' describing matrix and is made up of $p$ independent row vectors,
while $V$ is the `column effect' describing matrix and is made up of $q$ independent column vectors.
We further assume that the relationships between the variables
in each row vector of $U$ and those in each column vector of $V$
can be respectively explained by row hidden factors $F$ and column hidden factors $E$,\\
\centerline{$U_{i.} = LF_{i} + \eta_{i},i=1,...,p,$}
\centerline{$V_{.j} = \Lambda E_{j} + \xi_{j},j=1,...,q.$}
As a result, we call this model \textbf{two-way factor model} (2wFM),
because both row and column effects are described
by hidden factors.
Figure 1 (a) is a Bayesian network
representation for 2wFM. It can be seen that
2wFM is essentially a generalization of
classical factor models in analyzing matrix data.
It inherits the dimension-reduction idea from
 classical factor models and can directly distinguish between
 row and column effects by introducing a hidden group of factors
$F$ and $E$. Figure 1 (b) further illustrates the differences between
2wFM and classical factor models on one-way scale. In the 2wFM framework,
hidden factors affect the observable variables with a specific selection.

\begin{figure}[t]
  \centering
    \scalebox{0.4}{\includegraphics{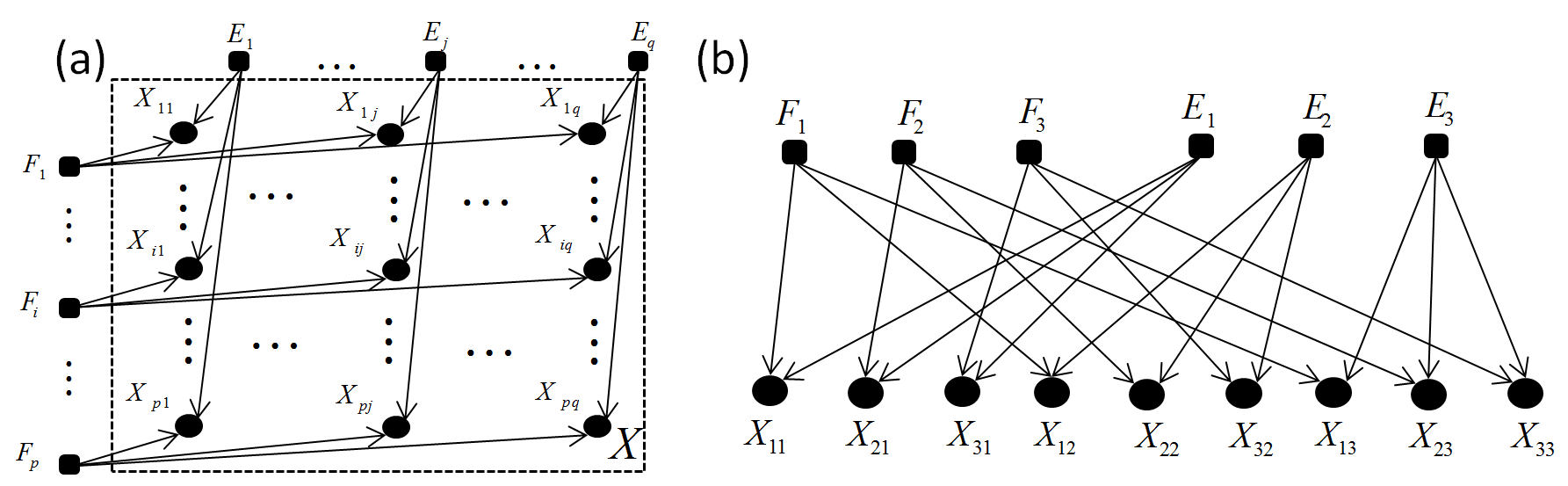}}\\
  \caption{ (a) A Bayesian network representation of 2wFM. (b) A bipartite graph representation of $\vec(X)$ when $p =3$ and $q = 3$.}\label{map_X}
  \normalsize
\end{figure}

Our contributions to the factor analysis of high-dimensional matrix-valued
data in this work exist in two parts. In the first we achieve
maximum likelihood estimation (MLE) of all parameters in the settings of 2wFM.
The specific structure in the covariance matrix of $\vec(X)$, denoted by $\Sigma_X$, brings difficulties in terms of
getting an analytical expression of the log-likelihood function.
Here, we generalize a conclusion introduced by \cite{Miller:1981} on calculating the inverse of
a special kind of matrix with `a nonsingular matrix plus a singular matrix' form and
obtain the exact form of $\Sigma^{-1}_X$ as well as $|\Sigma_X|$
under a group of identification conditions.
Due to the entanglements that exist between factor loadings and
the variance parameters of random factors and noises,
we further implement a block alternating maximizing
strategy to get the MLE for each parameter. The proposed algorithm
includes alternatively updating factor loadings and the
variance parameters.

The second part of our contributions refers to studying the theoretical properties of the MLE.
Under general conditions, we derive the consistency properties
as well as the asymptotic distribution, i.e., the central limit theorem,
for each estimator. As can be seen in Subsection 2.5, factor loadings of row and column factors
are combined with each other in the estimating equations, which makes it difficult
to study each of them separately. Moreover, without replication information from the original data set,
the convergence rates of many terms that constituted by the random factors and
the estimated factor loadings cannot be directly identified.
This fact has motivated us to undertake an in-depth study on the likelihood function and MLE.
The final results present a phenomenon in which the variance of the estimates
under the two-way factor model depends on the distance of variances
of row factors and column factors in a way that could not have been expected
in the classical factor analysis.
The asymptotic variance of row factor loadings is a trade-off between its own variance and
the variance of column factors (and vice versa).
The distance between the variances of row and column factors
has a heavy influence on the estimations of factor loadings,
a very small difference may result in large fluctuations in the factor loading estimations. On the other hand,
a small positive lower-bounded distance would lead to an effective asymptotic variance of
both row and column estimated factor loadings.

The rest of the paper is organized as follows. In Section 2, we mathematically describe and explain our model,
show the analytical expression of the likelihood function, analyze its basic structures, and
present our block alternating algorithm so as to obtain the MLE for each parameter. In Section 3,
we systematically discuss the main theoretical results for the
estimators. Results from simulations and real data analysis are given in Section 4.
All proof is displayed in the Appendix and supplementary material.
Throughout the paper, $A_{p\times q}$ denotes a $p \times q$ matrix.
In particular, $\textbf{0}_{q \times r}$ represents a $q\times r$ zero matrix, and
$\textbf{0}_{r}$ is a zero vector with $r$ entries. $\textbf{1}_{r}$ denotes a $r$ dimensional vector with each entry being equal to $1$,
$\textbf{1}_{r \times c}$ is a $r \times c$ matrix with each entry being
equal to $1$. $e_{k}^{(m)}$ represents a $k$
dimensional vector where the $m$th entry is 1 and other entries are 0.
$I_{k}=\big( e_{k}^{(1)}~e_{k}^{(2)}~\cdot\cdot\cdot~e_{k}^{(k)} \big)$
denotes the $k$ dimensional identity matrix.
$\delta_{ij}$ is a $\delta$ function. $||.||_2$ denotes the squared Euclidean distance
of a matrix. For simplicity, if $A$ is a $r$ dimensional vector and
there exists a $\delta>0$ such that each entry of $A$ can be
controlled by $p^{\delta}$ as $p \rightarrow \infty$, then $A$ is written as $A = O_{p}(p^{\delta})\textbf{1}_{r}$. If $A$ is a $r \times c$ matrix,
we write $A$ as $O_{p}(p^{\delta})\textbf{1}_{r \times c}$.

\csection{The two-way factor model}

Throughout this paper, we focus on covariance matrix analysis for high-dimensional
matrix-valued data and ignore the
mean-related parameter inference. We assume that the number of row and column factors are already known and
do not need to be estimated.
We first introduce our model in Subsection 2.1;
we then illustrate the MLE problem and the analytical expression of the
log-likelihood function in Subsection 2.2 and 2.3. Computation details
for obtaining the MLEs are discussed in Subsection 2.4. Subsection 2.5 concerns on the estimating equations of MLE.

\csubsection{The two-way factor model and its identification conditions}

In summary, our model can be directly written as
\begin{eqnarray*}\label{TWF4}
({\rm MC1}) & {\rm Additivity}~~~~ &    X = FL^{T} + \Lambda E^{T} + \epsilon, \nonumber \\
({\rm MC2}) &{\rm Independence}            &    F,~ E {\rm ~and ~}\epsilon~{\rm are~ mutually ~independent;} \nonumber \\
({\rm MC3}) & {\rm Factors~~~~~~~~}&    F = (F_{1\cdot}~  ...~ F_{p\cdot})^{T}\in \mathbb{R}^{p \times r}, E = (E_{1\cdot}~ ...~ E_{q\cdot})^{T} \in \mathbb{R}^{q \times c} \nonumber \\
            &             &    F_{1\cdot},...,F_{p\cdot}\stackrel{\text{i.i.d}}{\sim} N_{r}(0, \Psi_{F}), E_{1\cdot},...,E_{q\cdot}\stackrel{\text{i.i.d}}{\sim} N_{c}(0, \Psi_{E}); \nonumber \\
({\rm MC4}) &{\rm Noises~~~~~~~~~} &    \epsilon = [\epsilon_{ij}]_{p\times q},
                                       \epsilon_{11},\dots,\epsilon_{pq} \stackrel{\text{i.i.d}}{\sim} N(0, \sigma^2);   \nonumber \\
({\rm MC5}) &{\rm Parameters~~~}& \Psi_{F}= \diag(\sigma^{2}_{F_{1}},...,\sigma^{2}_{F_{r}}), \Psi_{E}= \diag(\sigma^{2}_{E_{1}},...,\sigma^{2}_{E_{c}}), \nonumber \\
            &                & \sigma^{2}_{F_{1}}>\cdots>\sigma^{2}_{F_r}>0, \sigma^{2}_{E_{1}}>\cdots>\sigma^{2}_{E_c}>0, \nonumber \\
            &                & \sigma^{2}_{F_{k}} \neq \sigma^{2}_{E_{m}}, {\rm for} ~ k=1,...,r, m=1,...,c.
 \end{eqnarray*}
Here $r$ and $c$ are the numbers of row and column factors,
$L\equiv(L_{1}\cdots L_{r}) \in \mathbb{R}^{q \times r}$ and $\Lambda\equiv(\Lambda_{1}\cdots\Lambda_{c}) \in \mathbb{R}^{p \times c}$
are the factor loadings for $F$ and $E$, and $L_{j}\in \mathbb{R}^{q},j=1,\dots,r,\Lambda_{i}\in\mathbb{R}^{p},i=1,\dots,c$.
To make the effects of each factor distinguishable, we make a common assumption, which has also appeared in several other works (\cite{Anderson:(1956)}, \cite{ Bai:2012}),
that the components of $F$ and $E$ have different variances, respectively, with an ordered
relationship in (MC5). Similar to classical factor model, 2wFM assumes that
the covariance structure between the entries of the data matrix $X$
can be effectively described by the linear combinations of several hidden factors.
When the components of $L$ or $\Lambda$ are all equal to zero, the above model would be reduced to the classical
factor model.

Under (MC1)-(MC5), the distribution of vectorized $X$ is
\begin{eqnarray}\label{TWF}
\vec(X) \sim N_{pq\times pq}\left(0, \Sigma_{X} \right)~\text{and}~\Sigma_{X}=I_{p} \otimes A + B \otimes I_{q}  + \sigma^{2}I_{p} \otimes I_{q},
\end{eqnarray}
where $A=L \Psi_{F} L^{T}$ is a $ q \times q $
symmetric matrix, and $B = \Lambda \Psi_{E} \Lambda^{T}$ is a $p \times p $ symmetric matrix.
It is natural to compare 2wFM to the well-known normal matrix variate model (MVM)
\begin{eqnarray*}\label{MVM}
X_{p\times q} \sim N_{pq} (M, \Phi_{p \times p} \otimes \Omega_{q \times q}),
\end{eqnarray*}
where $\Phi$ records the covariance of each column in $X$, and $\Omega$
records the covariance of each row in $X$, as mentioned in
\cite{Gupta:2000}, \cite{Adhikari:2007}, \cite{Allen:(2012)} and \cite{Zhou:2014}.
2wFM uses a different approach in decomposing the covariance
matrix of $X$. Not that provided by some kind of
matrix decomposition, such as direct production in MVM or Singular Value Decomposition (SVD) as proposed in (\cite{Wang Dong:2016}),
2wFM provides a direct separation to the row and column effects
by assuming an additive structure
existing between the row and
column factors and the random noises.
This setting induces
the covariance matrix $\Sigma_X$ consisting of (the sum of) three parts. The first part
is a full-ranked diagonal matrix corresponding to the noise
structure, and the second and third parts are Kronecker products of an
identity matrix with a low-rank matrix
representing the sparse effect induced by row and column latent factors.
This result can be further considered as a generalization to
the covariance structure obtained under classical factor model settings.

Without any further restrictions, parameters in ($\ref{TWF}$) cannot be directly identified.
This is a common problem in classical factor models that has been studied carefully by
\cite{Anderson:(1956)} and \cite{Bai:2012}.
In the context of our model, since there are hidden factors for both
row and column, this problem should be reconsidered.
In the following proposition, we present the identification conditions which should be held throughout this work.

\textbf{Identification Conditions}.
Let $\theta=(L,\Lambda,\Psi_{F},\Psi_{E},\sigma^{2})$ be all the parameters. For model ($\ref{TWF}$), there are two conditions \\
(IC1)  $\frac{L^TL}{q\sigma^2} = I_{r}$, $\frac{\Lambda^T\Lambda}{p\sigma^2} = I_{c}$;\\
(IC2)  Two factor loadings, such as $L$ and $L'$, are considered as equivalent if and only if there exists a diagonal matrix $D$ with diagonal
entries equal to 1 or $-1$ such that $L = L'D$.

\textbf{Proposition 1.}\label{P1}\emph{ Under (IC1) and (IC2), if there exists $\theta^{'}  = (L^{'}, \Lambda^{'}, \Psi^{'}_{F}, \Psi^{'}_{E}, \sigma^{'2})$ such that $\Sigma_{X}(\theta) = \Sigma_{X}(\theta^{'})$, then $\theta = \theta^{'}$.
}

(IC1) and (IC2) are two restrictions that are often imposed in factor analysis. Proposition 1 states that
by ignoring the effect of two directions on a line in (IC2),
it is enough for someone to impose restrictions on row and column factor loadings and the variance of random errors,
and no more restrictions need to be considered for the parameter identification problem in  model ($\ref{TWF}$).
In the next section, we will show that under (IC1) and (IC2), model (\ref{TWF}) not only has concise form on likelihood function but also
 provides a convenient way to update the parameters in the iterative process of obtaining the MLE.

\csubsection{The MLE problem}

For the model in (\ref{TWF}), the log-likelihood function can be written as
\begin{eqnarray} \label{logliksingle}
\ln\ell(\theta;X) \propto - \ln \left| \Sigma_{X} \right| - \vec(X)^{T}   \Sigma_{X}^{-1} \vec(X).
\end{eqnarray}
The maximum likelihood estimation for $\theta$ can then be defined as
\begin{eqnarray} \label{MLE1}
\hat{\theta} = \mathop{ \arg\max}_{\theta \in \Theta} \left\{  -\ln \left| \Sigma_{X} \right| - \vec(X)^{T}   \Sigma_{X}^{-1} \vec(X) \right\},
\end{eqnarray}
with $\Theta = \{\theta:\theta \text{ satisfies (MC5),~(IC1) and (IC2)} \}$.

($\ref{MLE1}$) is a difficult optimization problem because (a) the number of parameters contained
in $\theta$ is proportional to $p$ and $q$ and (b) the specific structure of $\Sigma_{X}$ makes the log-likelihood function
in ($\ref{MLE1}$) difficult to obtain. Although several methods have been suggested for when
this situation occurs \citep{Sheena:2003,Vandenberghe:2004,Wainwright:2006,Yuan:2007,Yuan:2009},
they are only designed to handle problems, and most of them essentially follow indirect ways to handle
the relevant optimization problems. Additionally, they often require more assumptions, such as sparsity,
on the structure of $\Sigma_{X}$ in order to obtain the theoretical properties
\citep{Sheena:2003,Wainwright:2006,Friedman:2007,Bickel:2008,Won:2013,Dahl:2008}.
In this work, we instead seek a more direct way to solve ($\ref{MLE1}$).
We strive to obtain the closed form of $\Sigma^{-1}_{X}$ and $|\Sigma_{X}|$,
then provide an analytical expression of ($\ref{logliksingle}$). As can be seen in Subsection 2.4,
this results in a direct strategy for solving the problem ($\ref{MLE1}$).

\csubsection{The analytical expression of the log-likelihood function}

$A$ and $B$ in $\Sigma_{X}$ can be written as
\begin{eqnarray*}
A &=& L\Psi_{F}L^T = \sigma_{F_{1}}^{2}L_{1}L_{1}^{T}+\cdots+\sigma_{F_{r}}^{2}L_{r}L_{r}^T \triangleq A_{1} + \cdots + A_{r}, \nonumber \\
B &=& \Lambda\Psi_{E}\Lambda^T = \sigma_{E_{1}}^{2}\Lambda_{1}\Lambda_{1}^{T} + \cdots + \sigma_{E_{c}}^{2}\Lambda_{c}\Lambda_{c}^T\triangleq B_1 + \cdots + B_{c},
\end{eqnarray*}
where $A_{j} = \sigma_{F_{j}}^{2}L_{j}L_{j}^{T}$, $j=1,\dots,r$, $B_{i}=\sigma_{E_{i}}^{2}\Lambda_{i}\Lambda_{i}^{T}$, $i=1,\dots,c$.
To get $\Sigma_{X}^{-1}$ and $|\Sigma_{X}|$, we generalize Miller's results (\cite{Miller:1981}) in the following proposition. \\
\textbf{Proposition 2.}\label{P3}\emph{
(a) Let $W = G\otimes I_{N} + I_{M} \otimes E$, where $E$ is an $N$-dimensional semi-positive definite matrix of rank $r$, and $G$ is an $M$-dimensional positive definite matrix. Then,
$W^{-1} = G^{-1}\otimes I_{N} - \sum_{i=1}^r  [(G+\lambda^E_{i}I_{M})^{-1}G^{-1}] \otimes E_i$;\\
(b) Let $W = I_{N}\otimes G + E \otimes I_{M}$. Then,
$W^{-1} = I_{N} \otimes G^{-1} - \sum^{r}_{i=1}E_i \otimes (G+\lambda^{E}_i I_{M})^{-1}G^{-1}$, \\
where $E_i = \lambda^E_{i} e_i e_i^T$, $\lambda^E_{i}$ and $e_i$ are respectively the $i$th eigenvalue and eigenvector of $E$ with
$ e_i^{T} e_j  = \delta_{ij}$, $\delta_{ij}$ being the delta function, is equal to 1 if $i=j$, otherwise, equal to 0.
}

In the case of there being single row and column factors, i.e., $r =1$ and $c = 1$,
$\Sigma_{X}$ can be simplified as
\begin{eqnarray*}
\Sigma_{X} &=&I_{p}\otimes A + B\otimes I_{q} + \sigma^{2}I_{p}\otimes I_{q}\\
&=&I_{p}\otimes \sigma^{2}_{F}LL^{T} + \sigma^{2}_{E}\Lambda\Lambda^{T}\otimes I_{q} + \sigma^{2}I_{p}\otimes I_{q}\\
 &=&(\sigma^{2}_{E}\Lambda\Lambda^{T} + \sigma^{2}I_{p}) \otimes I_{q} + I_{p} \otimes \sigma^{2}_{F}LL^{T}.
 \end{eqnarray*}
By (a) in Proposition 2, $\Sigma^{-1}_{X}$ can then be written as
\begin{eqnarray*}\label{Sigmainverse}
\Sigma_{X}^{-1} = d_{1} I_{p} \otimes I_{q} - d_{2} I_{p} \otimes A -d_{3} B \otimes I_{q} + d_{4} B \otimes A,
\end{eqnarray*}
where
$  d_1= \frac{1}{\sigma^{2}},~  d_2=\frac{1}{\sigma^{2}(\sigma^{2}+\sigma_{F}^{2} L^{T}L)},~ d_3=\frac{1}{\sigma^{2}(\sigma^{2}+\sigma_{E}^{2}\Lambda^{T}\Lambda)} $ and\\
$  d_4 = \frac{1}{\sigma^{2}_{F}\sigma^{2}_{E}L^{T}L\Lambda^{T}\Lambda}\left(\frac{1}{\sigma^{2}}-
\frac{1}{\sigma^{2}+\sigma^{2}_{F}L^{T}L} - \frac{1}{\sigma^{2}+\sigma^{2}_{E}\Lambda^{T}\Lambda}
+ \frac{1}{\sigma^{2}+\sigma^{2}_{F}L^{T}L+\sigma^{2}_{E}\Lambda^{T}\Lambda} \right). $

For the closed form of $|\Sigma_{X}|$, let $Q_{A}\widetilde{A}Q^{T}_{A}$ and $Q_{B}\widetilde{B}Q^{T}_{B}$
respectively be the eigen-decompositions of $A$ and $B+ \sigma^2I_{p}$, where $Q_{A}$ and $Q_{B}$ are orthogonal matrices,
$\widetilde{A}=\diag(\sigma^{2}_{F}L^{T}L,0,\dots,0)$, and $\widetilde{B} =\diag(\sigma^{2}_{E}\Lambda^{T}\Lambda+\sigma^{2},\sigma^{2},\dots,\sigma^{2})$ are diagonal matrices with entries in the principal diagonal equal to the eigenvalues of $A$ and $B+ \sigma^2I_{p}$. According to the property of the Kronecker product, $\Sigma_{X}$ can then be decomposed as\\
\centerline{$\Sigma_{X} = \left( Q_{B} \otimes Q_{A} \right) \left( I_{p} \otimes \widetilde{A} + \widetilde{B} \otimes I_{q} \right) \left( Q^{T}_{B} \otimes Q^{T}_{A} \right)$,}
hence the eigenvalues of $\Sigma_{X}$ can be identified easily from $I_{p} \otimes \widetilde{A} + \widetilde{B} \otimes I_{q}$. As a result,
$|\Sigma_{X}|$ can be written as
$$|\Sigma_{X}| = ( \sigma^2+ \sigma^{2}_{F}L^{T}L+ \sigma^{2}_{E}\Lambda^{T}\Lambda)( \sigma^2+ \sigma^{2}_{F}L^{T}L)^{p-1}
( \sigma^2+\sigma^{2}_{E}\Lambda^{T}\Lambda)^{q-1} (\sigma^{2})^{(p-1)(q-1)}. $$
When there are multiple row and column factors in 2wFM, we have\\
\textbf{Proposition 3.}\label{P3}\emph{
 In general case, i.e., $r \geq 1, c\geq 1$, with (IC1) and (IC2), \\
(a) $\Sigma_{X}^{-1} = \left( I_{p}\otimes A + B \otimes I_{q} + \sigma^{2}I_{p} \otimes I_{q} \right)^{-1}$ \\
$~~~~~~~~~~=d_{1}I_{p} \otimes I_{q} - \sum_{j=1}^{r}{d_{2j}I_{p}\otimes A_{j}} - \sum_{i=1}^{c}{d_{3i}B_{i}
\otimes I_{q}} + \sum_{j=1}^{r}\sum_{i=1}^{c}{d_{4ij}B_{i}\otimes A_{j}}, $\\
where
  $d_{1} = \frac{1}{\sigma^{2}}, ~ d_{2j}=\frac{1}{\sigma^{4}(1+q\sigma^{2}_{F_{j}})}, ~ d_{3i}=\frac{1}{\sigma^{4}(1+p\sigma^{2}_{E_{i}})}$ and\\
  $d_{4ij} = \frac{1}{pq\sigma^{6}\sigma^{2}_{F_{j}}\sigma^{2}_{E_{i}}}\Big( 1- \frac{1}{1+q\sigma^{2}_{F_{j}}}
 - \frac{1}{1+p\sigma^{2}_{E_{i}}} + \frac{1}{1+q\sigma^{2}_{F_{j}}+p\sigma^{2}_{E_{i}}} \Big).  $\\
(b) $\left|\Sigma_{X}\right| = \left( \sigma^{2} \right)^{pq}\prod_{j=1}^{r} \left( 1+q\sigma^{2}_{F_{j}} \right)^{p-c} \prod_{i=1}^{c}\left( 1+p\sigma^{2}_{E_{i}} \right)^{q-r} \prod_{i=1}^{c}\prod_{j=1}^{r}\left( 1+q\sigma^{2}_{F_{j}} + p\sigma^{2}_{E_{i}} \right).$
}

According to Proposition 3,
the log-likelihood function $(\ref{logliksingle})$ can be written as
\begin{eqnarray}\label{2.2.33}
\ln\ell\left( \theta;X \right) &\propto& -\sum^{c}_{i=1}\sum^{r}_{j=1}\ln \left( 1 + q\sigma^{2}_{F_{j}}  + p\sigma^{2}_{E_{i}} \right) -(p-c)\sum^{r}_{j=1} \ln\left( 1 + q\sigma^{2}_{F_{j}}  \right) \nonumber \\
&& - (q-r)\sum^{c}_{i=1}\ln\left( 1+p\sigma^{2}_{E_{i}}\right) - pq\ln\sigma^2 -d_{1}Q_{1}\nonumber \\
&& + \sum_{j=1}^{r}{\sigma^{2}_{F_{j}}d_{2j}Q_{2j}} + \sum_{i=1}^{c}{\sigma^{2}_{E_{i}}d_{3i}Q_{3i}} - \sum_{j=1}^{r}\sum_{i=1}^{c}{\sigma^{2}_{F_{j}}\sigma^{2}_{E_{i}}d_{4ij}Q_{4ij}},
\end{eqnarray}
where $Q_1=\tr(X^{T}X)$, $Q_{2j}=L_{j}^{T}X^{T}XL_{j}$, $Q_{3i}=\Lambda_{i}^{T}XX^{T}\Lambda_{i}$ and $Q_{4ij}=L_{j}^{T}X^{T}\Lambda_{i}\Lambda_{i}^{T}XL_{j}$.

\csubsection{Block alternating maximizing strategy for MLE}

Here, $\theta$ is split into three groups $L$, $\Lambda$, $(\Psi_F, \Psi_E, \sigma^2)$,
and a block alternating maximizing strategy is proposed
to calculate their MLE. This design induces to non-decreasing updates for the
value of the log-likelihood function ($\ref{2.2.33}$) (Proposition S1 in Subsection A.4 in the supplementary material).
More specifically, updating each parameter group proceeds as follows:
\begin{description}
  \item[1.] (Initialization) Initialize $L^{(0)},\Lambda^{(0)},\Psi_F^{(0)},\Psi_E^{(0)},\sigma^{2(0)}$ (the method for choosing the initial values can be referred to in Subsection A.2 in the supplementary material for more detail) and set
              $\text{err}_0=0.01$ and $\epsilon_{0}=0.005$.

  \item[2.]
  Given $L^{(m)},\Lambda^{(m)},\Psi_F^{(m)},\Psi_E^{(m)},\sigma^{2(m)}$, update $L$ to $\widetilde{L}^{(m+1)}$ by maximizing $$ \sum_{j=1}^{r} L_j^T W_{j}^{(m)}L_j  ~ \text{s.t.} ~  L^TL=qI_{r}. $$
 If $r=1$, then
 $\widetilde{L}^{(m+1)}=\sqrt{q}\nu_{\max}\big( W_{1}^{(m)} \big)$, where $\nu_{\max}(W_0)$ is the unit eigenvector corresponding to the largest eigenvalue of $W_0$.

 If $r>1$, then let $\lambda_{L}=\min_{j=1}^{r} \lambda_j^{\min}-\epsilon_{0},$
 where $\lambda_j^{\min}$ denotes the smallest eigenvalue of $W_j^{(m)}$, and let $A_j^{(m)}=-\lambda_{L} I_q+W_j^{(m)}$.
        Maximize  $ \sum_{j=1}^{r} L_j^T A_{j}^{(m)}L_j$ subject to $L^TL=I_{r}$ through the following iterative steps;
  \begin{description}
   \item[(2.1)] Initialize $L^{(m_0)}=L^{(m)}$ and $t=0$;
   \item[(2.2)] Let $A_L^{(m_t)}=(A_1^{(m)}L_1^{(m_t)}\cdots A_{r}^{(m)}L_{r}^{(m_t)})$. By SVD decomposition,
   we get $A_L^{(m_t)}=U_LD_LV_L^T$ ($U_{L},D_{L}$ and $V_{L}$ are $q\times r$, $r\times r$ and $r\times r$ matrices, respectively). Let $L^{(m_{t+1})}=U_LV_L^T$;
   \item[(2.3)] Let $f^{(t+1)}=\sum_{j=1}^{r} L_j^{(m_{t+1})T} A_{j}^{(m)}  L_j^{(m_{t+1})}$. If $|f^{(t+1)}-f^{(t)}|<\text{err}_{0}$, then $\widetilde{L}^{(m+1)}=\sqrt{q}L^{(m_{t+1})}$, else $t=t+1$ and repeat steps (2.2)-(2.3).
 \end{description}

  \item[3.]
  Given $\widetilde{L}^{(m+1)},\Lambda^{(m)},\Psi_F^{(m)},\Psi_E^{(m)},\sigma^{2(m)}$, update $\Lambda$ to $\widetilde{\Lambda}^{(m+1)}$ by maximizing
    $$ \sum_{i=1}^{c} \Lambda^T_i M_{i}^{(m)}\Lambda_i  ~ \text{s.t.} ~  \Lambda^T \Lambda=pI_{c}. $$
  If $c=1$, then
 $\widetilde{\Lambda}^{(m+1)}=\sqrt{p}\nu_{\max}\big( M_{1}^{(m)} \big)$.

 If $c>1$, let $\mu_{\Lambda}=\min_{i=1}^{c} \mu_i^{\min}-\epsilon_{0},$ where $\mu_i^{\min}$ denotes the smallest eigenvalue of $M_i^{(m)}$, and $B_i^{(m)}=-\mu_{\Lambda} I_p+M_i^{(m)}$.
        Maximize  $ \sum_{i=1}^{c} \Lambda_i^T B_{i}^{(m)}\Lambda_i$ subject to $\Lambda^T \Lambda=I_{c}$ through the following iterative steps;
   \begin{description}
   \item[(3.1)] Initialize $\Lambda^{(m_0)}=\Lambda^{(m)}$ and $t=0$;
   \item[(3.2)] Let $B_{\Lambda}^{(m_t)}=(B_1^{(m)}\Lambda_1^{(m_t)}\cdots B_{c}^{(m)}\Lambda_{c}^{(m_t)})$. By SVD, we can get $A_\Lambda^{(m_t)}=U_\Lambda D_\Lambda V_\Lambda^T$ ($U_{\Lambda},D_{\Lambda}$ and $V_{\Lambda}$ are $p\times c$, $c\times c$ and $c\times c$ matrices, respectively). Let $\Lambda^{(m_{t+1})}=U_\Lambda V_\Lambda^T$;
   \item[(3.3)] Let $g^{(t+1)}=\sum_{i=1}^{c} \Lambda_i^{(m_{t+1})T} B_{i}^{(m)}  \Lambda_i^{(m_{t+1})}$. If  $|g^{(t+1)}-g^{(t)}|<\text{err}_{0}$, then $\widetilde{\Lambda}^{(m+1)}=\sqrt{p}\Lambda^{(m_{t+1})}$, else $t=t+1$ and repeat steps (3.2)-(3.3).
 \end{description}

  \item[4.] Given $\widetilde{L}^{(m+1)},\widetilde{\Lambda}^{(m+1)},\Psi_F^{(m)},\Psi_E^{(m)},\sigma^{2(m)}$, update ($\Psi_F,\Psi_E,\sigma^2$) to ($\widetilde{\Psi}_F^{(m+1)},\widetilde{\Psi}_E^{(m+1)},\widetilde{\sigma}^{2(m+1)}$) with the EM algorithm:
      \begin{description}
        \item[(4.1)] Initialize $\Psi_F^{(m_0)}=\Psi_F^{(m)}, \Psi_E^{(m_0)}=\Psi_E^{(m)},\sigma^{2(m_0)}=\sigma^{2(m)} $ and $t=0$;
        \item[(4.2)] Update $( \Psi_F^{(m_t)},\Psi_E^{(m_t)},\sigma^{2(m_t)})$ to $(\widetilde{\Psi}_F^{(m_{t+1})},\widetilde{\Psi}_E^{(m_{t+1})},\widetilde{\sigma}^{2(m_{t+1})} )$ with the EM equations given in Subsection A.3 in the supplementary material;
        \item[(4.3)] Let $h^{(m_{t+1})}=\ln \ell(\widetilde{L}^{(m+1)},\widetilde{\Lambda}^{(m+1)},\Psi_F^{(m_{t+1})},\Psi_E^{(m_{t+1})},\sigma^{2(m_{t+1})})$.
            If $|h^{(m_{t+1})}-h^{(m_{t})}| < \text{err}_{0},$ then  $(\widetilde{\Psi}_F^{(m+1)},\widetilde{\Psi}_E^{(m+1)},\widetilde{\sigma}^{2(m+1)})
            =(\widetilde{\Psi}_F^{(m_{t+1})},\widetilde{\Psi}_E^{(m_{t+1})},\widetilde{\sigma}^{2(m_{t+1})})$, else $t=t+1$ and repeat steps (4.2)-(4.3).
      \end{description}

  \item[5.][Rotation] Let $U^{(m+1)}_{F}$ be the eigenvectors of $\frac{1}{\widetilde{\sigma}^{2(m+1)}}\widetilde{\Psi}_F^{(m+1)}$ \\
  and $D_F^{(m+1)}=U_{F}^{(m+1)}(\frac{1}{\widetilde{\sigma}^{2(m+1)}}\widetilde{\Psi}_F^{(m+1)})U^{(m+1)T}_{F}$. Let $V^{(m+1)}_{E}$ be the eigenvectors of $\frac{1}{\widetilde{\sigma}^{2(m+1)}}\widetilde{\Psi}_E^{(m+1)}$ and $D_E^{(m+1)}=V^{(m+1)}_{E}(\frac{1}{\widetilde{\sigma}^{2(m+1)}}\widetilde{\Psi}_E^{(m+1)})V^{(m+1)T}_{E}$. Set
      \begin{eqnarray*}
        \sigma^{2(m+1)} &=& \widetilde{\sigma}^{2(m+1)},~ \Psi_F^{(m+1)} = D_F^{(m+1)},~\Psi_E^{(m+1)} = D_E^{(m+1)}, \nonumber \\
        L^{(m+1)} &=& \widetilde{L}^{(m+1)}\widetilde{\Psi}_F^{(m+1)\frac{1}{2}}U^{(m+1)T}_{F}D_F^{(m+1)-\frac{1}{2}},  \nonumber \\
        \Lambda^{(m+1)} &=& \widetilde{\Lambda}^{(m+1)}\widetilde{\Psi}_E^{(m+1)\frac{1}{2}}V^{(m+1)T}_{E}D_E^{(m+1)-\frac{1}{2}}. \nonumber
      \end{eqnarray*}
  \item[6.] Let $ln^{(m+1)}=\ln \ell(L^{(m+1)},\Lambda^{(m+1)},\Psi_F^{(m+1)},\Psi_E^{(m+1)},\sigma^{2(m+1)})$.
            If $|ln^{(m+1)}-ln^{(m)}| < \text{err}_{0},$ then $(\hat{L},\hat{\Lambda},\hat{\Psi}_F,\hat{\Psi}_E,\hat{\sigma}^{2})= (L^{(m+1)},\Lambda^{(m+1)},\Psi_F^{(m+1)},\Psi_E^{(m+1)},\sigma^{2(m+1)})$, else $m=m+1$ and repeat steps 2-6.
 \end{description}
Remarks. \\
1. Given $\Lambda, (\Psi_{F},\Psi_{E},\sigma^2)$ and $L^TL = q\sigma^2I_{r}$, ($\ref{2.2.33}$) can be written as
\begin{eqnarray} \label{2.7}
\ln\ell(\theta;X)  &\propto& \sum_{j=1}^{r}{\sigma^{2}_{F_{j}}d_{2j}L^{T}_{j}X^{T}XL_{j}} - \sum_{j=1}^{r}\sum_{i=1}^{c}{\sigma^{2}_{F_{j}}\sigma^{2}_{E_{i}}d_{4ij}L^{T}_{j}X^{T}\Lambda_{i}\Lambda^{T}_{i}XL_{j}}  \nonumber \\
&=& \sum_{j=1}^{r} L^{T}_{j} W^{L}_{j} L_{j},
\end{eqnarray}
where $W^{L}_{j} = \sigma^{2}_{F_{j}}X^{T}[d_{2j}I_{p} - \sum_{i=1}^{c}{\sigma^{2}_{E_{i}}d_{4ij}\Lambda_{i}\Lambda^{T}_{i}}]X$,
$j = 1, ... , r$. When $r = 1$, ($\ref{2.7}$) can be simplified as
\begin{eqnarray}\label{2.24}
\ln\ell(\theta;X)  \propto \sigma_{F}^{2} L^{T}\left( d_2X^{T}X - \sigma^{2}_{E}d_4X^{T}\Lambda \Lambda^{T} X \right)L,
 \end{eqnarray}
which is just a quadratic form of $L$. $(\ref{2.24})$ is maximized when
$$ L = \sqrt{q\sigma^2} \cdot \nu_{\max} \left( d_2X^{T}X-\sigma^{2}_{E}d_4X^{T}\Lambda \Lambda^{T} X \right). $$
When $r \geq 1$,
($\ref{2.7}$) is the sum of a series of quadratic form as follows
\begin{eqnarray} \label{fx}
g(L) =  \sum^{r}_{j=1} L^{T}_{j} W^{L}_{j} L_{j} ~  \text{s.t.}  ~ L^{T}_{i}L_{j} = q\sigma^{2}\delta_{ij} \ (1 \leq i,j \leq r  ),
\end{eqnarray}
where $W^{L}_{1}, \dots, W_{r}^{L}$ are positive semidefinite $q\times q$ matrices.
It is well known that an analytical solution can be obtained for the problem of maximizing $g(L)$
when $W^{L}_1 = \cdots = W^{L}_r$. However, orthogonal restrictions on each pair of $(L_{i},L_{j})$ $i,j=1,\dots,r$,
with distinct structures of $W^{L}_{1},\dots,W^{L}_{r}$ lead to difficulties in
obtaining the solutions of maximizing ($\ref{fx}$).
Here, two previous works \citep{Bolla:(1998), Bolla:(2001)} are introduced to overcome
these difficulties.
\cite{Bolla:(1998)} and \cite{Bolla:(2001)} indicate that
for any solution of (\ref{fx}),
there exists a $r\times r$ symmetric matrix $A_{L}$ such that
this solution must satisfy
$$(W_1^{L}L_1\cdots W^{L}_rL_r)=LA_{L}.$$
Thus, an iterative updating process can be implemented in
three steps,
(a) assembling $W_1^{L}L_1,\dots, W^{L}_rL_r$ to a new matrix $\widetilde{W}_{L}=(W_1^{L}L_1\cdots W^{L}_rL_r)$, (b) obtaining
SVD to $\widetilde{W}_{L}$
and (c) getting an update of $L$. \cite{Bolla:(1998)} and \cite{Bolla:(2001)} proved
that this process can make the value of the object function $g$ converge to a local maximum.\\
2. Symmetrically, given $L, (\Psi_{F},\Psi_{E},\sigma^2)$ and $\Lambda^T\Lambda = p\sigma^2I_{c}$, ($\ref{2.2.33}$) can be written as
\begin{eqnarray} \label{2.8}
\ln\ell(\theta;X)  &\propto& \sum_{i=1}^{c}{\sigma^{2}_{E_{i}}d_{3i}\Lambda^{T}_{i}XX^{T}\Lambda_{i}} - \sum_{j=1}^{r}\sum_{i=1}^{c}{\sigma^{2}_{F_{j}}\sigma^{2}_{E_{i}}d_{4ij}\Lambda^{T}_{i}XL_{j}L^{T}_{j}X^{T}\Lambda_{i}}  \nonumber \\
& = &  \sum_{i=1}^{c} \Lambda^{T}_{i} W^{\Lambda}_{i}   \Lambda_{i},
\end{eqnarray}
where $W^{\Lambda}_{i} = \sigma^{2}_{E_{i}}X[d_{3i}I_{q} - \sum_{j=1}^{r}{\sigma^{2}_{F_{j}}d_{4ij}L_{j}L^{T}_{j}}]X^{T}$,
for $i = 1, ..., c$.\\
3. Given $L$ and $\Lambda$, updating $(\Psi_{F},\Psi_{E},\sigma^{2})$.
As opposed to $(\ref{2.7})$ and $(\ref{2.8})$, there are generally no analytical solutions to updating $(\Psi_{F},\Psi_{E},\sigma^{2})$.
Here, the EM method are adopted to update $\Psi_F$, $\Psi_E$ and $\sigma^2$ in order to definitely be able to obtain a local optimal solution. The details can be referred to Subsection A.3 in the supplementary material.

\csubsection{Estimating equations of $\hat{\theta}$}

In this section, we study the estimating equations
of $\hat{\theta}$ \citep{Anderson:(1956), Bai:2012}.
Although there are restrictions that (IC1) and (IC2) impose on $\hat{\theta}$,
we can prove (using Lagrange multiplier techniques) that the properties of $\hat{\theta}$ that satisfy
$$\hat{\theta} =  \mathop{ \arg\max}_{\theta \in \Theta} \ln\ell(\theta; X) ~ \text{ s.t.}~ \dfrac{L^TL}{q\sigma^2} = I_{r}, ~ \dfrac{\Lambda^T\Lambda}{p\sigma^2} = I_{c},$$
can be studied based on the estimating equations with (IC1) and (IC2):
\begin{eqnarray}
&\dfrac{\partial \ln\ell}{\partial \theta} \Big|_{\theta = \hat{\theta}} = \textbf{0}, ~ \dfrac{\hat{L}^{T}\hat{L}}{q\hat{\sigma}^2} = I_{r}, ~ \dfrac{\hat{\Lambda}^{T}\hat{\Lambda}}{p\hat{\sigma}^2} = I_{c}.
\end{eqnarray}
This means that the above equations include all the information from $\hat{\theta}$
in terms of discovering the relationships between each parameter.
A detailed expression of each estimating equation is presented in the supplementary material (Subsection A.8).

\csection{Asymptotic properties of MLE}

In this section, the statistical properties of the MLE
in a large sample framework are discussed. It is important to note that the number
of parameters in our model, such as parameters contained in factor loadings $L$ and $\Lambda$,
increases as $p$ and $q$ diverge. This fact broadly exists in high-dimensional
data analysis and has been becoming a growing concern in recent theoretical studies.
As illustrated by \cite{Bai:2012}, methods that originate from the Taylor expansion
such as the delta method cannot be applied directly to this situation.
This is mainly because that the tail terms, which could be ignored in classical fixed or lower dimensional factor models, become the sum of infinity terms and each of them is $o_{p}(1)$ as $p$ and $q$ diverge. Thus, it is difficult to bound the tail terms in proper convergence orders.

Rather than approximating the distance between
the true values of parameters $\theta^{*}$ and their estimations $\hat{\theta}$ by using
the variations of the linear part of the likelihood function valued in the local area of $\hat{\theta}$,
as is done in the Taylor expansion, we instead follow the technical route proposed by \cite{Bai:2012}.
Broadly speaking,
we look for a direct algebraic decomposition of the log-likelihood function ($\ref{2.2.33}$)
such that the rates of each term obtained from the decomposition can be identified through technical
analysis. However, due to the complexity of ($\ref{2.2.33}$) as well as
there being no replication information from the original data set,
this is an extremely difficult task.

Before introducing our theoretical results in detail, we shall first form the following asymptotical conditions:\\
(AC1) $p/q \rightarrow y \in (0,\infty)$;\\
(AC2) $\left\|L_{m\cdot}\right\|_{2}=o(p^{0.5})$, $m=1,\dots,q$ and $\left\|\Lambda_{k\cdot}\right\|_{2}=o(p^{0.5})$, $k=1,\dots,p$;\\
(AC3) There exists a large enough positive constant $C$, such that \\$\left(\diag^{T}{(\Psi_{F})},\diag^{T}{(\Psi_{E})},\sigma^2\right)^{T} \in [C^{-1}, C]^{r+c+1}.$

Remarks. Conventional studies often impose assumptions on the relationships
between the number of variables, $p$, and the sample size, $n$,
while in matrix-valued data analysis, special attention must be paid to
the size of the data set $X$. (AC1) sets $p$ and $q$ with an equal speed of divergence.
We think that this restriction is natural and reasonable
because if the diverging order of $p$, for example, is larger than $q$, then the parameters in the
factor loadings $\Lambda$ will increase far faster than those in $L$, and
the information brought about by new introduced data may not be enough to
distinguish and measure the effects of the row and column factors. On the other hand,
inspired by the case in the classical factor model,
people usually make assumptions that the order of $p$ is no larger than the order of $n$
\citep{Bai:2012, Fan:(2008)}.
We follow this method and find that this can induce a consistency result
for each estimator. (AC2) and (AC3) are
similar to (C.1) and (C.2) listed in \cite{Bai:2012}. (AC2) is a generalized version of (C.1) in which
each row entry of factor loadings $L$ and $\Lambda$ is $O(1)$.

\textbf{Proposition 4.}\label{P4}\emph{
Under the model conditions (MC1)-(MC5), identification conditions (IC1)-(IC2), and asymptotic conditions (AC1)-(AC3), let $\left( \hat{\Psi}_{F}, \hat{\Psi}_{E}, \hat{\sigma}^{2}\right)$ be the maximum likelihood estimation of $\left( \Psi_{F},\Psi_{E},\sigma^{2}\right)$ in likelihood function $(\ref{logliksingle})$.
There then exists a large enough constant $\tilde{C}$,
such that $\left( \diag^{T}(\hat{\Psi}_{F}), \diag^{T}(\hat{\Psi}_{E}), \hat{\sigma}^{2}\right)^{T} \in \left[ \tilde{C}^{-1},\tilde{C} \right]^{r+c+1}$ $in ~probability$.
}

Proposition 4 presents the boundedness property for $( \hat{\Psi}_{F}, \hat{\Psi}_{E}, \hat{\sigma}^{2})$.
This condition is a basic result for deriving the large sample properties of the estimators, although it has often appeared as an underlying assumption
in previous work \citep{Bai:2012,Doz:(2012)}.
Along with Theorem 1, a rigorous proof of boundedness has been given here.
We compare the marginal values of the log-likelihood
function ($\ref{logliksingle}$) when $\Psi_{F}$, $\Psi_{E}$ and $\sigma^{2}$ individually approach to $0^{+}$ and $+\infty$, and to
the value when $\theta$ locates at its true value $\theta^*$. We find that the values of $\ln \ell(\theta)/p$
and $\ln \ell(\theta)/q$ at $\theta^{*}$
are bounded in probability and will tend to $-\infty$ when $\Psi_{F},\Psi_{E}$ and $\sigma^{2}$ approach to $0^{+}$ or $+\infty$,
thus verifying the conclusion in Proposition 4.

\textbf{Theorem 1 (Consistency).}\label{T1}\emph{
Under the model conditions (MC1)-(MC5), identification conditions (IC1)-(IC2), and asymptotic conditions (AC1)-(AC3), as $p,q \rightarrow \infty$, we have\\
$(a)$ in simple case that $r =1,c = 1$,
\begin{eqnarray*}\label{3.8}
\hat{\sigma}^{2}_{F} - \sigma^{*2}_{F} \stackrel{p}{\longrightarrow}  0 ,~~~
\hat{\sigma}^{2}_{E} - \sigma^{*2}_{E} \stackrel{p}{\longrightarrow}  0 ,~~~
\hat{\sigma}^{2} - \sigma^{*2} \stackrel{p}{\longrightarrow}  0,\\
\dfrac{(\hat{L}-L^{*})^{T}(\hat{L}-L^{*})}{q\hat{\sigma}^2} \stackrel{p}{\longrightarrow} 0 ,~
\dfrac{(\hat{\Lambda}-\Lambda^{*})^{T}(\hat{\Lambda}-\Lambda^{*})}{p\hat{\sigma}^2} \stackrel{p}{\longrightarrow} 0.
\end{eqnarray*}
$(b)$ In general case that $r \geq 1, c\geq 1$,
\begin{eqnarray*}
  \diag\left( \hat{\Psi}_{F} - \Psi^{*}_{F} \right) \stackrel{p}{\longrightarrow}  0 ,~
\diag\left( \hat{\Psi}_{E} - \Psi^{*}_{E} \right) \stackrel{p}{\longrightarrow}  0 ,~
 \hat{\sigma}^{2} - \sigma^{*2}  \stackrel{p}{\longrightarrow}  0,\\
tr\left[ \dfrac{(\hat{L}-L^{*})^{T}(\hat{L}-L^{*})}{q\hat{\sigma}^2} \right] \stackrel{p}{\longrightarrow} 0 ,~
tr\left[ \dfrac{(\hat{\Lambda}-\Lambda^{*})^{T}(\hat{\Lambda}-\Lambda^{*})}{p\hat{\sigma}^2} \right] \stackrel{p}{\longrightarrow} 0.
\end{eqnarray*}
}
Here, the idea of `average convergence' is taken from \cite{Bai:2012} and \cite{Doz:(2012)} for the
consistency expression of the factor loading estimators ($\hat{L}, \hat{\Lambda}$)
since there will be infinity estimators for $L$ and $\Lambda$ as
$p$ and $q$ diverge. This is a crucial step in the traditional M-method (in \cite{Bai:2012}) in proving
$ \mathop{\sup}_{\theta \in \Theta} |R(\theta)| = o_p(1),$
where $R(\theta)$ is a remainder obtained after decomposing
the original likelihood function $\ln\ell(\theta; X)$ into two parts
$\ln\ell(\theta; X) = \ln\widetilde{\ell}(\theta; X) + R(\theta),$
where $\ln\widetilde{\ell}(\theta; X)$ is maximized at the true value $\theta^*$;
therefore, if $\mathop{\sup}_{\theta \in \Theta} |R(\theta)| = o_p(1)$,
it means that $\hat{\theta}$ and $\theta^*$ can asymptotically be closed in some sense.
However, in 2wFM, due to the fact that there are no replicates for $X$ and
the temporary lack of the boundedness property of $\hat{\theta}$,
this conclusion for $R(\theta)$ cannot be obtained.
Therefore, more investigations into $\ln\ell(\theta; X)$ shall be undertaken.

The process of getting the consistency results is divided into three steps.
In the first step, we focus on ($\hat{L}$, $\hat{\Lambda}$). A thorough study on the MLE is performed,
which are induced by a degenerate form of ($\ref{MLE1}$) and (\ref{2.2.33}), that is,
\begin{eqnarray}\label{sop}
\left( \hat{L}^{(\uppercase\expandafter{\romannumeral1})},\hat{\Lambda}^{(\uppercase\expandafter{\romannumeral1})} \right) = \mathop{\arg\max}_{ L^{T}L =q\sigma^{*2}, \Lambda^{T}\Lambda=p\sigma^{*2}} \left( \dfrac{L^{T}X^{T}XL}{pq^{2}\sigma^{*4}} + \dfrac{\Lambda^{T}XX^{T}\Lambda}{p^{2}q\sigma^{*4}} - \dfrac{\Lambda^{T}XLL^{T}X^{T}\Lambda}{p^{2}q^{2}\sigma^{*6}} \right).
\end{eqnarray}
As an optimization problem first appeared in Lemma 6A,
solutions to ($\ref{sop}$), ($\hat{L}^{(\uppercase\expandafter{\romannumeral1})},
\hat{\Lambda}^{(\uppercase\expandafter{\romannumeral1})}$),
are easy to obtain and have close connections
to the real MLE ($\hat{L}, \hat{\Lambda}$). Relevant conclusions
on ($\hat{L}^{(\uppercase\expandafter{\romannumeral1})},
\hat{\Lambda}^{(\uppercase\expandafter{\romannumeral1})}$) can be fully
generalized to the case of $(\hat{L}, \hat{\Lambda})$ (Lemma 8).
Based on the results regarding ($\hat{L}, \hat{\Lambda}$) from the first step and
the order estimations of some basic terms, in the second step
we go back to the original
likelihood function ($\ref{logliksingle}$), construct connections
between ($\hat{L}, \hat{\Lambda}$) and ($\hat{\Psi}_{F},
\hat{\Psi}_{E}, \hat{\sigma}^{2}$), and conclude that (Lemma 7)
$$\hat{\sigma}^{2}_{F} = \dfrac{\hat{L}^{T}X^{T}X\hat{L}}{pq^{2}\hat{\sigma}^{4}}+o_{p}(1),~
\hat{\sigma}^{2}_{E} = \dfrac{\hat{\Lambda}^{T}XX^{T}\hat{\Lambda}}{p^{2}q\hat{\sigma}^{4}}+o_{p}(1),~
\dfrac{\hat{L}^{T}X^{T}\hat{\Lambda}\hat{\Lambda}^{T}X\hat{L}}{p^{2}q^{2}\hat{\sigma}^{6}}= o_{p}(1).$$
In the third step, the terms in these relations are connected to their true value
$\theta^*$ (Lemma 9) and are substituted into the
estimating equation of $\hat{\sigma}^2$, the consistency property of
each estimator is then straightforward. \\
Remarks. \\
1. Here, only the main results obtained from
$r =1$ and $c = 1$ are explained. In general case, $r \geq 1, c \geq 1$, all terms can be
structured into fixed dimensional matrix form and the proof follows the same course.
Lemma 6B, Lemma 7, Lemma 8 and Lemma 9 can be referred to for more details.\\
2. The key problem during the process of obtaining the consistency results for each estimator
is to confirm the orders of the four terms
$$\dfrac{L^{*T}\hat{L}}{q\hat{\sigma}^2}, ~\dfrac{\Lambda^{*T}\hat{\Lambda}}{p\hat{\sigma}^2},~
\dfrac{E^T\hat{L}}{q\hat{\sigma}^2}, ~\dfrac{F^T\hat{\Lambda}}{p\hat{\sigma}^2}.$$
The first two can be regarded as a projection of the estimated factor loadings in the space
constituted by their true values, their order estimations also being the key problem in one-way
high-dimensional factor models \citep{Bai:2012}.
The final two terms can be regarded as a projection of
the estimated factor loadings in the space constituted by row and column factors.
In the classical high-dimensional factor model framework, only
$\frac{L^{*T}\hat{L}}{q\hat{\sigma}^2}$ needs to be dealt with, while in the context of our model,
it is necessary to simultaneously
estimate the orders of these four terms, which makes our problem more general and more of a challenge.
The optimization function in ($\ref{sop}$) is essentially a function of four highly similar terms
$$\dfrac{L^{*T}\hat{L}}{q\sigma^{*2}},~ \dfrac{\Lambda^{*T}\hat{\Lambda}}{p\sigma^{*2}}, ~\dfrac{E^T\hat{L}}{q\sigma^{*2}}, ~ \dfrac{F^T\hat{\Lambda}}{p\sigma^{*2}},$$
which therefore obtains the essential structure of ($\ref{sop}$).
As a result, the following conclusions can be drawn:
$$\dfrac{L^{*T}\hat{L}}{q\hat{\sigma}^2} = 1+o_p(1),~ \dfrac{\Lambda^{*T}\hat{\Lambda}}{p\hat{\sigma}^2} = 1 + o_p(1),~
\dfrac{E^T\hat{L}}{q\hat{\sigma}^2} = o_p(1),~ \dfrac{F^T\hat{\Lambda}}{p\hat{\sigma}^2} = o_p(1).$$
This means the space comprised of $L^*$ or $\Lambda^*$, respectively, keeps
the most part of $\hat{L}$ and $\hat{\Lambda}$, which reveals a fact that
we can deal with them separately.

\textbf{Theorem 2 (Central Limit Theorem).}\label{T2}\emph{
Under the model conditions (MC1)-(MC5), identification conditions (IC1)-(IC2) and asymptotic conditions (AC1)-(AC3), as $p,q \rightarrow \infty$, we have\\
$(a)$ in simple case that $r =1$ and $c = 1$,
\begin{eqnarray*}
\sqrt{p} \left( \hat{L}_{m\cdot}-L_{m\cdot}^{*} \right)    &\stackrel{d}{\longrightarrow}& N \left( 0, \dfrac{\sigma^{*2}}{\sigma^{*2}_F} + \dfrac{\sigma^{*2}\sigma^{*2}_{E}(y\sigma^{*2}_{E} + \sigma^{*2}_{F})}{(\sigma^{*2}_{F} - \sigma^{*2}_{E})^2} \right),~m=1,\dots,q, \\
  \sqrt{q} \left( \hat{\Lambda}_{k\cdot}-\Lambda_{k\cdot}^{*} \right)  &\stackrel{d}{\longrightarrow}& N \left( 0, \dfrac{\sigma^{*2}}{\sigma^{*2}_E} + \dfrac{\sigma^{*2}\sigma^{*2}_{F}(y\sigma^{*2}_{E} + \sigma^{*2}_{F})}{y(\sigma^{*2}_{F} - \sigma^{*2}_{E})^2} \right),~ k=1,\dots,p, \\
  \sqrt{p} \left( \hat{\sigma}_{F}^{2}-\sigma_{F}^{*2} \right)  &\stackrel{d}{\longrightarrow}& N \left( 0, 2\sigma^{*4}_{F} \right),\\
  \sqrt{q} \left( \hat{\sigma}_{E}^{2}-\sigma_{E}^{*2} \right)&\stackrel{d}{\longrightarrow}& N \left( 0, 2\sigma^{*4}_{E} \right), \\
  \sqrt{pq} \left( \hat{\tilde{\sigma}}^2-\sigma^{*2} \right) &\stackrel{d}{\longrightarrow}& N \left( 0, 2\sigma^{*4} \right),~\hat{\tilde{\sigma}}^{2} = \left( 1 + \dfrac{1}{p} + \dfrac{1}{q} \right)\hat{\sigma}^{2}.
\end{eqnarray*}
$(b)$ In general case that $r \geq 1, c\geq 1$,
\begin{eqnarray*}
  \sqrt{p} \left( \hat{L}_{m\cdot}-L_{m\cdot}^{*} \right) &\stackrel{d}{\longrightarrow}& N _{r}\left( 0, \Sigma_{L} \right),~m=1,\dots,q, \\
  \sqrt{q} \left( \hat{\Lambda}_{k\cdot}-\Lambda_{k\cdot}^{*} \right) &\stackrel{d}{\longrightarrow}& N_{c}\left( 0, \Sigma_{\Lambda}\right),~k=1,\dots,p, \\
  \sqrt{p}~ diag\left( \hat{\Psi}_{F}-\Psi_{F}^{*} \right) &\stackrel{d}{\longrightarrow}& N_{r} \left( 0, 2\left(\Psi_{F}^{*} \right)^2 \right), \\
  \sqrt{q}~ diag\left( \hat{\Psi}_{E}-\Psi_{E}^{*} \right) &\stackrel{d}{\longrightarrow}& N_{c} \left( 0, 2\left(\Psi_{E}^{*} \right)^2 \right), \\
  \sqrt{pq} \left(\hat{\tilde{\sigma}}^2-\sigma^{*2}\right) &\stackrel{d}{\longrightarrow}& N \left(0, 2\sigma^{*4}\right),~\hat{\tilde{\sigma}}^{2}= \left( 1 + \dfrac{c}{p} + \dfrac{r}{q} \right)\hat{\sigma}^{2},
\end{eqnarray*}
where
\begin{eqnarray*}
\Sigma_{L} &=& \sigma^{*2}\Psi_{F}^{*-1} + \diag\left(\sum_{i=1}^{c}{\dfrac{\sigma^{*2}\sigma^{*2}_{E_{i}}(y\sigma^{*2}_{E_{i}} + \sigma^{*2}_{F_{1}})}{(\sigma^{*2}_{F_{1}} - \sigma^{*2}_{E_{i}})^2}} ,\dots, \sum_{i=1}^{c}\dfrac{\sigma^{*2}\sigma^{*2}_{E_{i}}(y\sigma^{*2}_{E_{i}} + \sigma^{*2}_{F_{r}})}{(\sigma^{*2}_{F_{r}} - \sigma^{*2}_{E_{i}})^2}  \right), \nonumber \\
  \Sigma_{\Lambda} &=& \sigma^{*2}\Psi_{E}^{*-1} + \diag\left(\sum_{j=1}^{r}{\dfrac{\sigma^{*2}\sigma^{*2}_{F_{j}}(\sigma^{*2}_{F_{j}} + y\sigma^{*2}_{E_{1}})}{y(\sigma^{*2}_{E_{1}} - \sigma^{*2}_{F_{j}})^2}} ,\dots, \sum_{j=1}^{r}\dfrac{\sigma^{*2}\sigma^{*2}_{F_{j}}(\sigma^{*2}_{F_{j}} + y\sigma^{*2}_{E_{c}})}{y(\sigma^{*2}_{E_{c}} - \sigma^{*2}_{F_{j}})^2}  \right).
\end{eqnarray*}
}
Theorem 2 presents the behavior of each estimator in a large sample scenario
when both row and column factors exist.
Results for $\hat{\sigma}^2_F $ and $ \hat{\sigma}^2_E$ are similar to the results from \cite{Bai:2012}.
The size of $\sigma^{*2}_F$ or $\sigma^{*2}_E$ is positive correlation with the asymptotic variance of $\hat{\sigma}^{2}_{F}$
or $\hat{\sigma}^{2}_{E}$.
Results on $\hat{\sigma}^{2}$ in Theorem 2 show that $\hat{\sigma}^2$ is an asymptotic unbiased estimation
for $\sigma^{*2}$. Different from the results obtained from the classical factor model,
the variances of row and column factors have interaction effects on the behaviors of
$\hat{L}$ and $\hat{\Lambda}$. In detail,
the asymptotic variance for each entry of $\hat{L}$ when $r =1$ and $ c = 1$
can be written as
\begin{equation} \label{asyvar}
\sigma^{*2} \left[ \dfrac{1}{\sigma^{*2}_{F}} + \dfrac{1 + y\Delta}{(\Delta - 1)^2} \right],
\end{equation}
where $\Delta = \frac{\sigma^{*2}_{F}}{\sigma^{*2}_{E}}$.
Intuitively,
\begin{figure}[t]
  \centering
    \scalebox{0.5}{\includegraphics{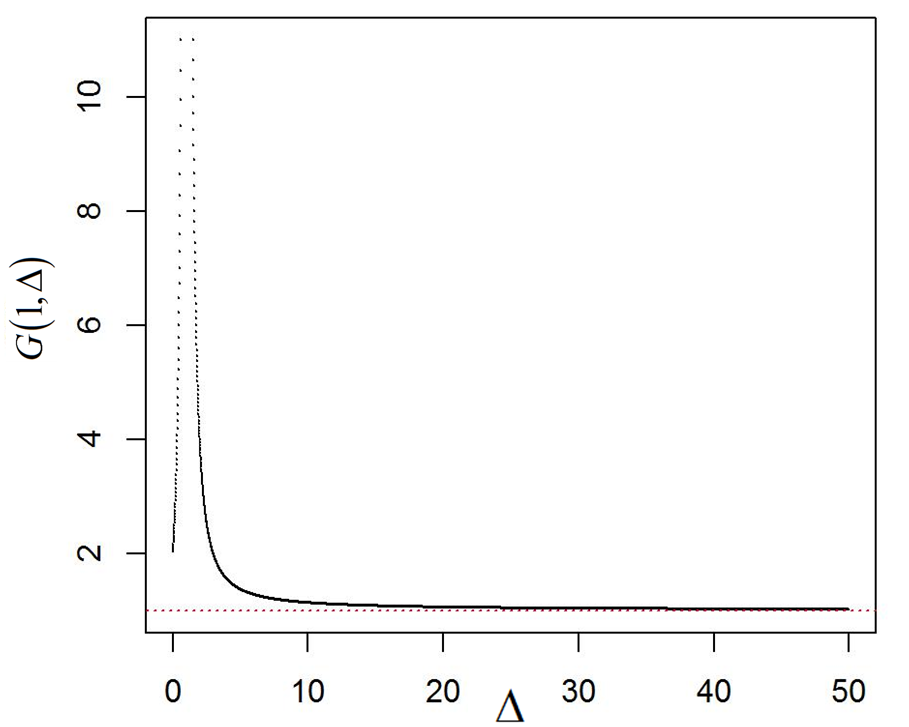}}\\
  \caption{\small Scatter plot for ($\ref{asyvar}$), labeled as G(1, $\Delta$), when $\sigma^{*2}=1$, $\sigma^{*2}_{F}=1$, $y=1$.
   $\Delta$ varies from 0 to 50. The red dotted line represents its limiting value.}\label{fig_Gdelta}
  \normalsize
\end{figure}
a small $\sigma^{*2}$ indicates a small disturbance from noises, while
a large $\sigma^{*2}_{F}$ lets the value of $F$ be distributed
across a wide range, thus bringing more useful information
on the inference of  $L$. Therefore, a small $\sigma^{*2}$
or a large $\sigma^{*2}_{F}$ will lead to a small asymptotic
variance of $\hat{L}$. The expression from (\ref{asyvar})
agrees well with this judgement.
Term $\frac{1 + y\Delta}{(\Delta - 1)^2}$ in (\ref{asyvar})
represents the degree of impact from the alternative column factor $E$.
Given $\sigma^{*2}$, $\sigma^{*2}_{F}$ and $y$, the size of (\ref{asyvar}) relates to
the value of $\Delta$, the ratio of $\sigma^{*2}_{F}$ and $\sigma^{*2}_{E}$.
It can be seen that a small distance between $\sigma^{*2}_{F}$ and $\sigma^{*2}_{E}$
will result in a large value of (\ref{asyvar}). Furthermore,
(\ref{asyvar}) may be very large when $\Delta$ is close to 1
(Figure 2
shows more details for the case in which $\sigma^{*2}= y=\sigma^{*2}_{F}=1$).
As a result, it
may need an extremely large number of $p$ and $q$ in order to make each entry of $\hat{L}$
be consistent with its true value. On the other hand,
a large distance between $\sigma^{*2}_{F}$ and $\sigma^{*2}_{E}$,
especially when $\sigma^{*2}_{F}=1$, $\sigma^{*2}_{E}\rightarrow 0^{+}$,
will make $\Delta$ close to 0, and the size of (\ref{asyvar}) in this situation
will coincide with the conclusion derived from the same settings of the classical factor model.
Similar conclusions can be obtained for $\hat{\Lambda}$ and for $(\hat{L},\hat{\Lambda})$
in general case $(r \geq 1, c\geq 1)$.\\
Remarks.\\
1. In order to obtain the results in Theorem 2, we need to deal with estimating equations which
report the restrictions that $\hat{\theta}$
must satisfy, and contain terms with important information on reflecting the limiting
behaviors of each parameter in $\hat{\theta}$. However, these terms often remain together and appear in several equations
as a series of cliques, and
as a result, their orders cannot be identified directly. This difficulty motivate us to find
a group of basic elements that connect these terms.
It is done by sufficiently
utilizing the relationships that exist in the estimating equations and substituting
these relationships into the estimating equations of $L$ and $\Lambda$.
The whole process is cumbersome with constant decomposing of each term,
calculating of the orders of new appeared terms, and keeping of the terms with higher
or unknown orders while unifying
the terms with known lower orders into one single term represented
by their common upper bound. This process continues
until there are no more elemental terms to be discovered.
Finally, two terms, $\frac{E^{T}(\hat{L} - L^{*})}{q\hat{\sigma}^{2}}$ and
$\frac{F^{T}(\hat{\Lambda} - \Lambda^{*})}{p\hat{\sigma}^{2}}$, come out to the surface
and it can be concluded that
all the other terms are essentially the functions of these two terms (Lemma 12).
We obtain their limiting distributions by
solving a group of two linear equations that are
induced during the simplifying process of the estimating equations
of $L$ and $\Lambda$ (Lemma 13). Results for other terms are then straightforward.

2. Identifying the orders of $\frac{\hat{L}^T(\hat{L} - L^{*})}{q\hat{\sigma}^{2}}$
and $\frac{\hat{\Lambda}^{T}(\hat{\Lambda} - \Lambda^{*})}{p\hat{\sigma}^{2}}$
is also a key step in getting the limiting distributions of each $\hat{\theta}$.
As opposed to the results of \cite{Bai:2012} which show that the order of
$\frac{\hat{L}^{T}(\hat{L} - L^{*})}{q\hat{\sigma}^{2}}$ is a lower one
compared to the order of $\hat{\sigma}^2$, we find that this term has
the same order as $\hat{\sigma}^{2} - \sigma^{*2}$ in our model settings.

\csection{NUMERICAL STUDIES}

\csubsection{Synthetic Data}

In this section, we implement four groups of simulations.
In the first and second simulation, we evaluate the accuracy of our
estimators under single and multiple row and column factors, respectively.
In the third simulation, we study the precision of $\hat{L}$ and $\hat{\Lambda}$
as the ratio of $\sigma^2_F$ and $\sigma^2_E$ varies across a wide range.
In the fourth simulation, we set $F$ and $E$ to follow Chi-square distributions and to test the robustness
performance of our method.

\begin{figure}[!htp]
\subfigure[$(p,q)=(50,50)$]{
\includegraphics[width=.3 \textwidth]{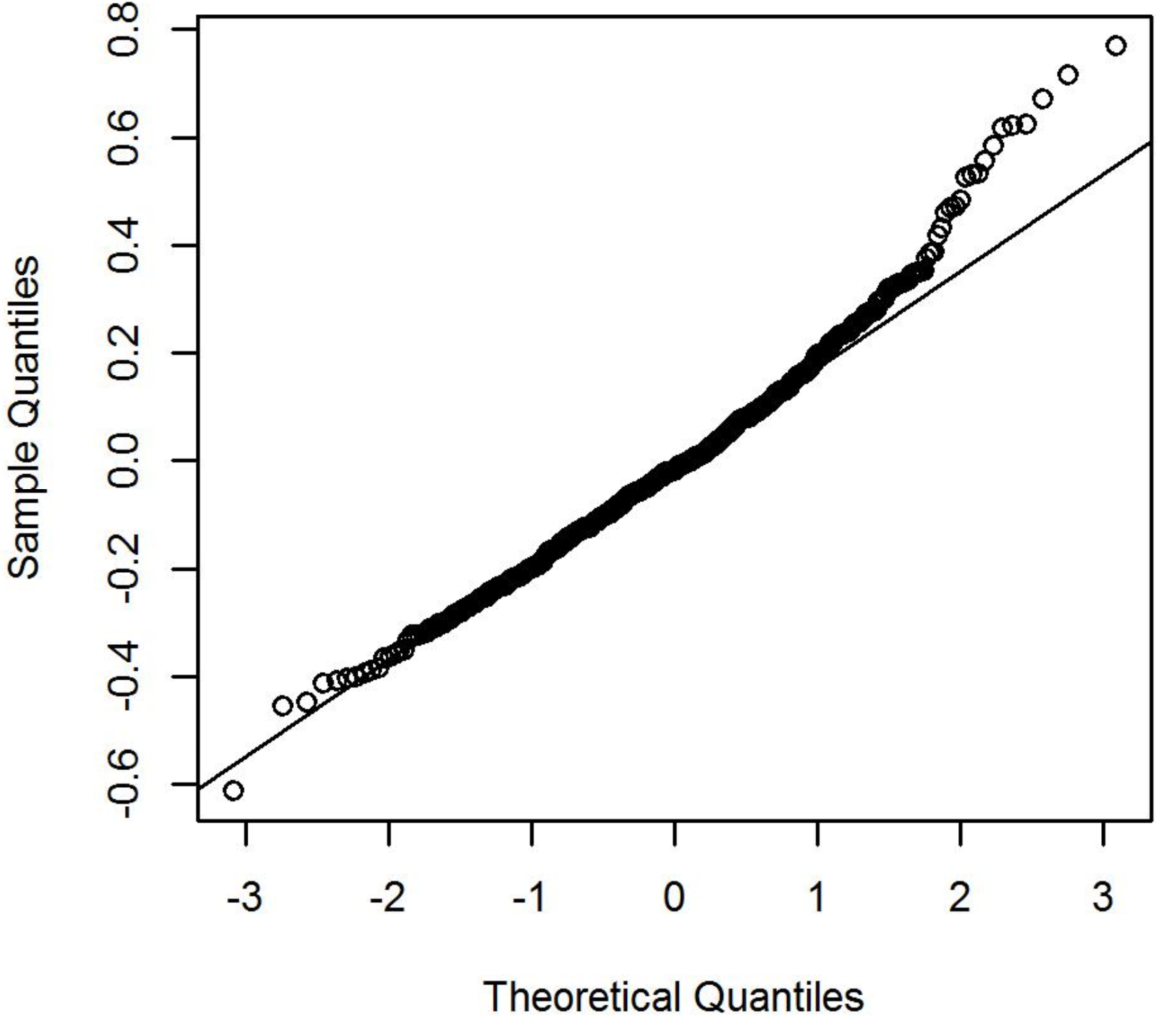}
}
\subfigure[$(p,q)=(200,200)$]{
\includegraphics[width=.3 \textwidth]{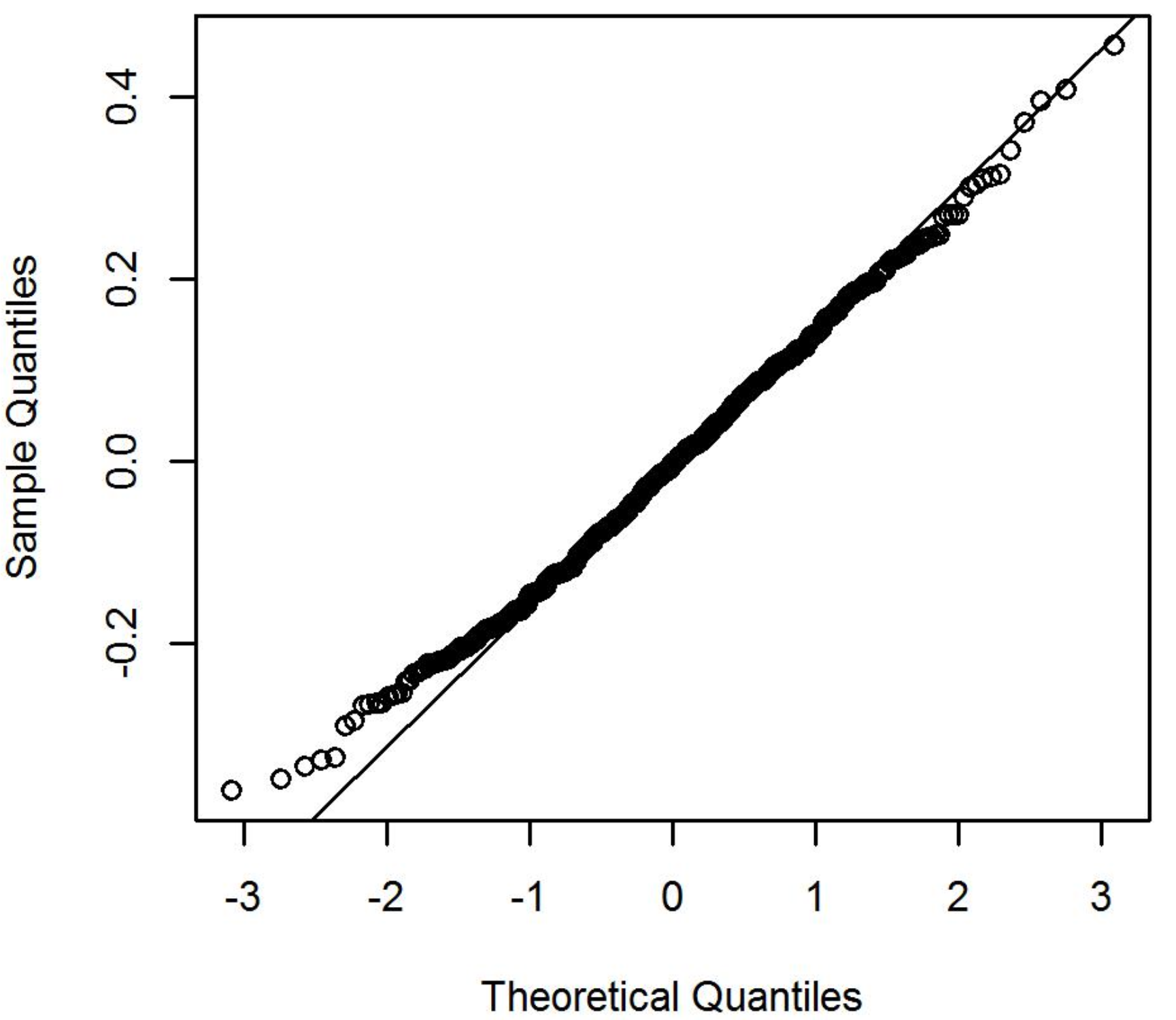}
}
\subfigure[$(p,q)=(1000,1000)$]{
\includegraphics[width=.3 \textwidth]{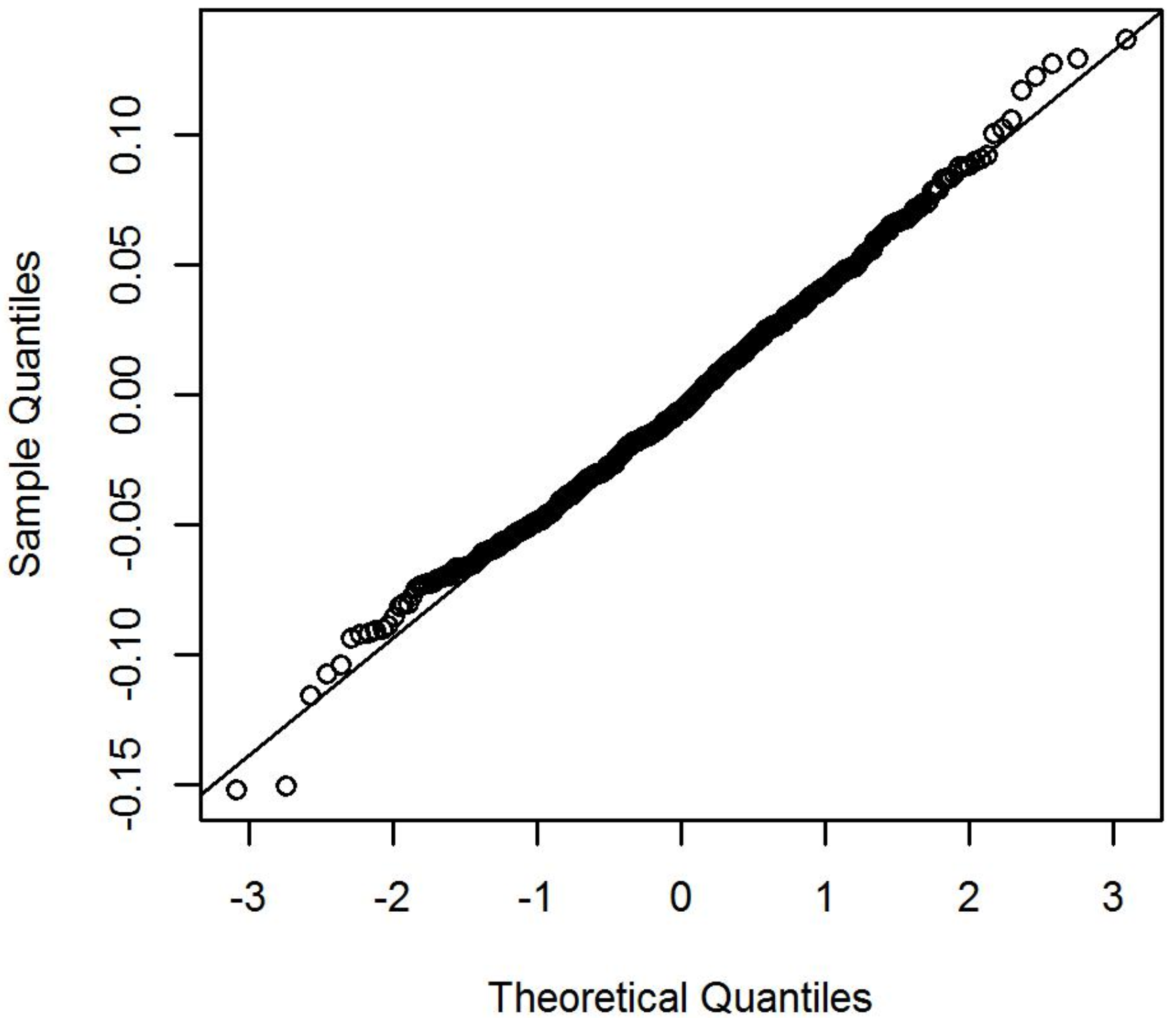}
}
\caption{Q-Q plot of ${\bf\hat{\Psi}}_E$ with ${\bf\Psi}_F=8$ in the accuracy study}\label{fig_E1}
\end{figure}

In the accuracy study,
we first set $r = 1$, $c = 1$,
$p =\{50, 100, 200, 500, 1000\}$, $q =\{50, 100, 200, 500, 1000\}$
and $\Psi_F = \{ 1.5, 2, 4, 8\}$.
For each pair of $p$, $q$ and each $\Psi_F$, we set
$\Psi_E= 1$, $\sigma^2 = 0.01$ (the same below) and draw 1000 samples (the same below) from the model ($\ref{TWF}$).
The elements of the factor loadings are first independently sampled from
uniform distribution $U[0, 1]$ and then normalized by the squared norms in order to
make the resulting factor loadings satisfy the identification condition (IC1). People could refer to Subsection A.1 in supplementary material for the sampling process when all parameters are given. MLE are achieved according to the algorithm in Subsection 2.4.
Table 1 and 2 only present the estimation results
for $p = \{50, 200, 1000\}$ and $q = \{50, 200, 1000\}$,
 the complete results are collected in Table S1 and S2 (in Subsection A.5).
Since the number of parameters in $L$ and $\Lambda$ are proportional to $q$ and $p$, we fit the estimated factor
loadings $\hat{L}$ and $\hat{\Lambda}$ to their true values and calculate the average $R^2$ as a measurement
of the accuracies(the same below). For $\hat{\Psi}_F$, $\hat{\Psi}_E$ and $\hat{\sigma}^{2}$, we
list the average value of each estimator with their mean absolute error (MAE) and mean square error (MSE) (the same below).
In Table 1 and 2,
there are two numbers in each grid. The first is the value of the MAE and the second one in brackets is
the value of the MSE. The magnitudes of the MAE and MSE for $\hat{\sigma}^2$ are respectively $10^{-4}$ and $10^{-7}$ (the same below).
It can be seen that the precisions of $\hat{L}$ and $\hat{\Lambda}$ are, respectively,
closely related to the size of $q$ and $p$. For each pair of $p$ and $q$,
$\hat{L}$ and $\hat{\Lambda}$ become more precise as $\Psi_{F}$
varied from $1.5$ to $8$.
$\hat{\sigma}^2$ converges to its true value more quickly
than $\hat{\Psi}_F$ and $\hat{\Psi}_E$, whereas the value changes of $\Psi_F$ and $\Psi_E$
have little impact on
the results for $\hat{\sigma}^2$. As $p$ and $q$ diverge, all estimators tend to
approach their true values, and these results are consistent with the conclusions from Theorem 1.
We also provide Q-Q plots in order to
give more details on the behaviors of each estimator.
 Figure \ref{fig_E1} provides Q-Q plots to give more details on the behaviors of ${\bf \hat{\Psi}}_E$. 
A noticeable improvement can be found as $p$ and $q$ increase.
We also obtain similar results for other estimators
(see Figure S2, S3, S4 and S5 in Subsection A.5 in the supplementary material for more details).

\begin{figure}[t]
  \centering
    \scalebox{0.7}{\includegraphics{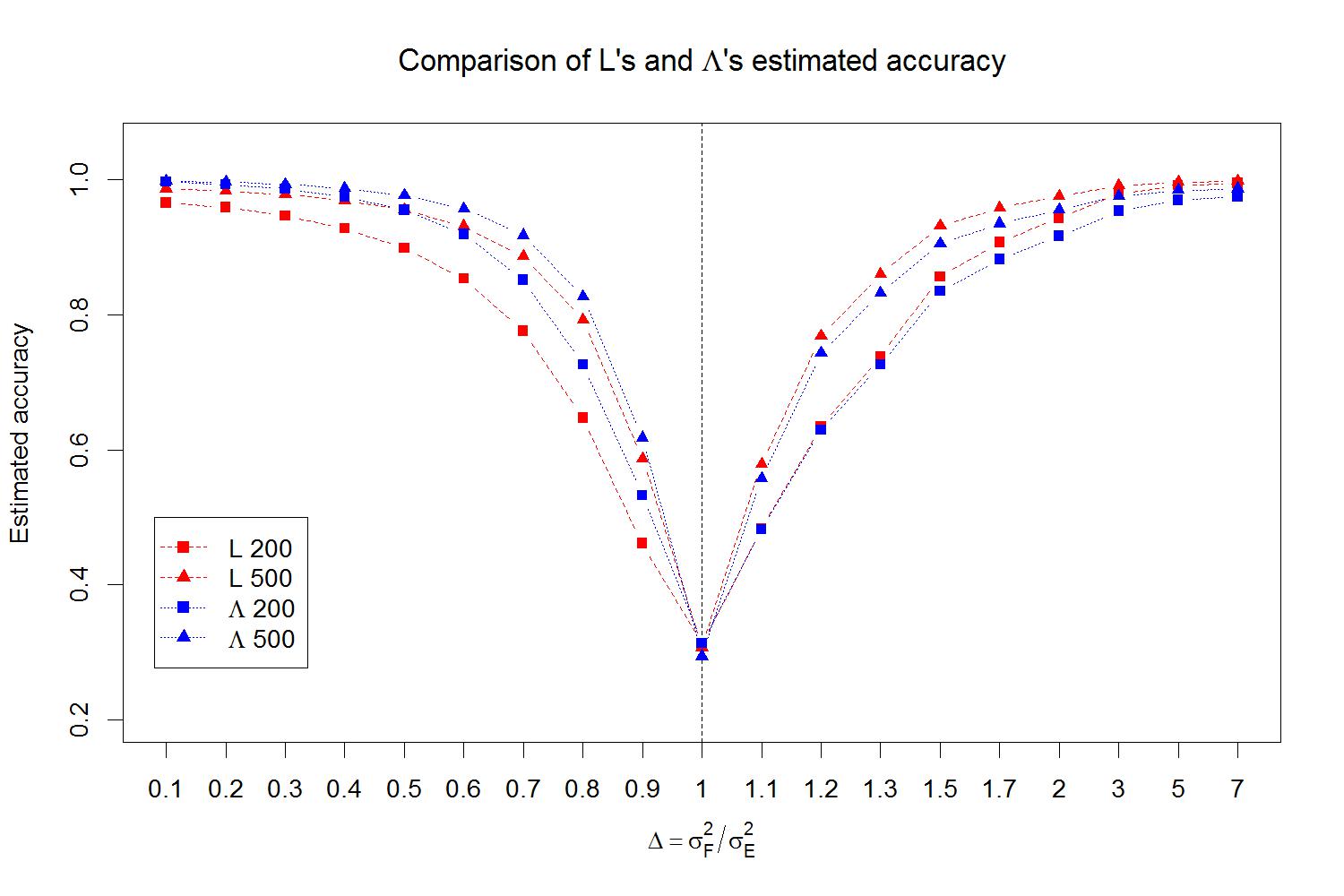}}\\
  \caption{\small Estimation accuracy comparison of $L$ and $\Lambda$.
  The dimensions of $L$ and $\Lambda$ are set to 200 and 500.
  Ratios of $\Psi_{F}$ and $\Psi_{E}$ vary from 0.1 to 7.}\label{fig_delta}
  \normalsize
\end{figure}

In the second simulation, we focus on the situation with multiple row and column factors.
At this time, $r$ is set to 2, $c$ is set to 3, and $\Psi_F$ and $\Psi_E$ are set to $\diag(10,8)$ and
$\diag(6,4,2)$, respectively. All other parameters, $p$, $q$ and $\sigma^2$ are set to the same values
used in the first simulation.
The results in Tables S3 and S4 (Subsection A.6) show that the number of factors, regardless of the rows
or columns, have inverse effects on the precision of $\hat{L}$ and $\hat{\Lambda}$,
and this phenomenon coincides with the conclusion for the asymptotic variance of $\hat{L}$ and $\hat{\Lambda}$ in Theorem 2.
On the other hand, this
effect can be omitted if $r$ and $c$ are not too large compared to the sizes of $p$ and $q$.

In the third simulation, we set $r =1$, $c = 1$, $\Psi_E  =4$,
$p =q\in\{ 200, 500\}$ and let $\Delta = \frac{\Psi_F}{\Psi_E} = \{0.1, 0.2, 0.3, 0.4, 0.5, 0.6, 0.7,
0.8, 0.9, 1, 1.1, 1.2, 1.3, 1.5, 1.7, 2, 3, 5, 7 \}$. For each pair $(\Delta,p)$, we derive the MLE, obtain the average estimated accuracy of $\hat{L}$ and $\hat{\Lambda}$, then present the two accuracy curves in Figure \ref{fig_delta}.  The curves indicate that
the distance of the values of $\sigma^2_F$ and $\sigma^2_E$ has a serious impact
on the precision of $\hat{L}$ and $\hat{\Lambda}$. With the same $p$ or $q$,
the asymptotic variances of $\hat{L}$ and $\hat{\Lambda}$ become less as
the ratio of $\sigma^2_F$ and $\sigma^2_E$ beyond 1 and as $p$ or $q$ increases, the accuracy of $\hat{L}$ and $\hat{\Lambda}$ are more close to 1. Those facts confirm the conclusions from Theorem 2. Another interesting phenomenon, which also appears in the first simulation, is that, when $\Delta<1$, the accuracy curves of $\hat{\Lambda}$ are superior to these of $\hat{L}$ (and vice versa).

To evaluate the robustness of our method, in the fourth simulation
we set $r =1$, $c = 1$, $p =\{200, 300, 400\}$,
$q =\{200, 300, 400\}$ and assume that factors $E$ and $F$ are generated
from a Chi-square distribution with both means equal to 0 and
$\Psi_E = 1$ and $\Psi_F = \{1.5, 2, 4\}$, respectively.
 The relative results are collected in Table S5 and S6 (see Subsection A.7 in the supplementary material).
Our method performs well for each parameter and can estimate
parameters with high precision even when common factors both $F$ and $E$ with non-normal distribution.

\newpage
\begin{landscape}
\begin{table}
\caption{Estimation Accuracy of Factor Loadings and $\sigma^2$ ($r=1$ and $c=1$)}
\label{Table1-S1}
\tabcolsep 2.0pt
\scalebox{0.90}{
\begin{tabular}{ccccccccccccccccccccccccccccccccc}
\hline
 & &\multicolumn{4}{c}{$L$}  & & \multicolumn{4}{c}{$\Lambda$}  & & \multicolumn{4}{c}{$\sigma^2$}  \vspace{0.15cm} \\
   \cmidrule(r){4-7}\cmidrule(r){9-12} \cmidrule(r){14-17}
$q$ & $p$  && 8 & 4 & 2 & 1.5 & & 8 & 4 & 2 & 1.5 & & 8 & 4 & 2 & 1.5 \\	
\hline

    &50  && 0.9819 & 0.9525 & 0.8300 & 0.6696  && 0.8871 & 0.8526 & 0.7335 & 0.5897 && 19.80(55.17) & 19.80(55.16) & 19.80(55.15) &19.80(55.14)\\
50  &200 && 0.9835 & 0.9581 & 0.8558 & 0.7067  && 0.9093 & 0.8834 & 0.7888 & 0.6555 && 11.79(18.66) & 11.79(18.66) & 11.78(18.66) & 11.78(18.66) \\
    &1000&& 0.9850 & 0.9619 & 0.8680 & 0.7251  && 0.9076 & 0.8809 & 0.7855 & 0.6511 && 9.759(10.75) & 9.758(10.75) & 9.758(10.75) & 9.758(10.75)\\
&&&&&&&&&&&\\
    &50  &&  0.9944 & 0.9810 & 0.8990 & 0.7660 && 0.9541 & 0.9285 & 0.8290 & 0.6968 && 11.76(18.78) & 11.76(18.78) & 11.75(18.78) & 11.75(18.78) \\
200 &200 &&  0.9957 & 0.9876 & 0.9429 & 0.8549 && 0.9748 & 0.9637 & 0.9156 & 0.8336 && 5.014(3.582) & 5.014(3.582) & 5.014(3.582) & 5.014(3.581)\\
    &1000&&  0.9965 & 0.9906 & 0.9609 & 0.8994 && 0.9760 & 0.9675 & 0.9316 & 0.8670 && 3.057(1.158) & 3.057(1.158) & 3.057(1.158) & 3.057(1.158) \\
&&&&&&&&&&&\\
    &50  &&  0.9972 & 0.9874 & 0.9166 & 0.7909 && 0.9740 & 0.9530 & 0.8679 & 0.7475 && 9.773(10.73) & 9.773(10.73) & 9.773(10.73) & 9.773(10.73) \\
1000&200 &&  0.9987 & 0.9951 & 0.9693 & 0.9068 && 0.9898 & 0.9830 & 0.9504 & 0.8884 && 3.047(1.156) & 3.047(1.156) & 3.047(1.156) & 3.047(1.156)  \\
    &1000&&  0.9991 & 0.9975 & 0.9876 & 0.9624 && 0.9937 & 0.9908 & 0.9772 & 0.9485 && 1.046(0.1485) & 1.046(0.1485) & 1.046(0.1485) & 1.046(0.1485) \\
\hline
\end{tabular}
}
\end{table}

\begin{table}
\caption{Estimation Accuracy of $\Psi_F$ and $\Psi_E$ ($r=1$ and $c=1$)}
\label{Table2}
 \tabcolsep 2.0pt
 \scalebox{0.90}{
\begin{tabular}{cccccccccccccccccccccccccccccccccccccc}
\hline
 & &\multicolumn{4}{c}{$\Psi_F$}  & & \multicolumn{4}{c}{$\Psi_E$}  \vspace{0.15cm} \\
   \cmidrule(r){4-7} \cmidrule(r){9-12}
$p$ & $q$  && 8 & 4 & 2 & 1.5  & & 8 & 4 & 2 & 1.5 \\	
\hline
    &50  && 1.2906(0.1975) & 0.6510(0.1346) & 0.3362(0.1190) & 0.2626(0.1377) &&0.1635(0.0420) & 0.1641(0.0419) & 0.1643(0.0413) & 0.1636(0.0407) \\
50  &200 && 1.2755(0.0988) & 0.6380(0.0692) & 0.3188(0.0663) & 0.2370(0.0811) &&0.0806(0.0098) & 0.0820(0.0102) & 0.0877(0.0118) & 0.0969(0.0146) \\
    &1000&& 1.2688(0.0788) & 0.6317(0.0575) & 0.3112(0.0583) & 0.2291(0.0740) &&0.0447(0.0032) & 0.0471(0.0037) & 0.0561(0.0057) & 0.0704(0.0091) \\
&&&&&&&&&&&\\
    &50  && 0.6518(0.1304) & 0.3310(0.0866) & 0.1754(0.0758) & 0.1459(0.0906) &&0.1662(0.0429) & 0.1651(0.0423) & 0.1623(0.0405) & 0.1566(0.0371) \\
200 &200 && 0.6372(0.0400) & 0.3183(0.0291) & 0.1591(0.0284) & 0.1210(0.0352) &&0.0805(0.0101) & 0.0804(0.0100) & 0.0804(0.0100) & 0.0812(0.0101) \\
    &1000&& 0.6373(0.0144) & 0.3190(0.0125) & 0.1599(0.0147) & 0.1205(0.0197) &&0.0364(0.0021) & 0.0367(0.0021) & 0.0379(0.0022) & 0.0400(0.0025) \\
&&&&&&&&&&&\\
    &50  && 0.3197(0.1208) & 0.1683(0.0777) & 0.1018(0.0674) & 0.0997(0.0802) &&0.1656(0.0433) & 0.1650(0.0428) & 0.1628(0.0409) & 0.1579(0.0376) \\
1000&200 && 0.3005(0.0338) & 0.1511(0.0222) & 0.0779(0.0195) & 0.0621(0.0235) &&0.0811(0.0100) & 0.0809(0.0100) & 0.0804(0.0098) & 0.0795(0.0095) \\
    &1000&& 0.2970(0.0115) & 0.1486(0.0078) & 0.0746(0.0069) & 0.0561(0.0084) &&0.0362(0.0020) & 0.0362(0.0020) & 0.0362(0.0020) & 0.0363(0.0020) \\
\hline
\end{tabular}
}
\end{table}
\end{landscape}

\csubsection{Real Examples}

In the context of city air quality assessment,
the standard method for measuring air quality is to calculate
the air quality index (AQI) according to the volumes of several
monitored pollutants, such as sulfur dioxide ($SO_2$) and
nitrogen dioxide ($NO_2$). However, AQI only reports
the maximum readings (linear transformed) of all the
pollutants
and does not consider the geographical relationships between cities.
Here, our method is applied in order to give a new explanation of the air quality for
each city.

Data are obtained from China National Environmental Monitoring Center website (http://www.cnemc.cn/).
We selected a typical data set $X$ containing 338 cities and 14 air pollutant indices.
Figure 5(A) presents a heatmap showing the whole data set which
reveals possible correlations existing
between cities as well as between pollutants.
We centralize $X$ by columns, set $r =1$ and $c = 1$ and apply our method to
estimate $F_i, i=1,\dots,338$ for each city and $E_j, j=1,\dots, 14$
for each pollutant. Figure 5(B) compares the AQIs with $\hat{F}_i, i=1,\dots,338$
in a scatterplot, which shows a high coincidence with the relative $R^2$ being as large as 0.895.
This result indicates that $F_i, i=1,\dots,338,$
can be regarded as measurements of the air qualities.
In order to investigate the impact of geographical factors, we apply K-means and classify $\hat{F}_i, i=1,\dots,338,$
into 6 clusters. The relevant results are collected in Figure 5(C), in which
cities in the same cluster are labeled with the same color.
Several areas, such as those located around
Shandong and Henan province, with less-favorable air qualities can be identified intuitively.
Figure 5(C) further indicates that our method can integrate geographic location information
and can thus provide better results than AQI does.
For example, Xi'an and Xianyang are two very close cities and are divided into
different air quality levels by the AQI-based method, while our method considers them to be at the same level.

As well as illustrating the air quality level for each city, our method further
provides an effective evaluation for each pollutant. Here, a new quantity, $R^2$, is calculated,
for each pollutant, which is the $R^2$ of the AQI regressed on the readings
of 14 pollutants. Obviously, pollutants with large $R^2$ are probably important
components of bad air quality. Figure 5(D) compares
$\hat{E}_j$ with $R^2_j$, $j=1,\dots,14$. All pollutants are intuitively
classified into 3 clusters. The six pollutants with the largest
absolute values of $\hat{E}_j$ have the smallest $R^2_j$, whereas two pollutants
(PM2.5 and PM10) with the smallest $\hat{E}_j$ (near zero) have the largest $R^2_j$.
This means that $E_j, j=1,\dots, 14$ can be considered as measurements
that quantify the degree of each pollutant in the composition of the air pollution.

\begin{figure}[t]
  \centering
    \scalebox{0.55}{\includegraphics[width = 1.8\textwidth]{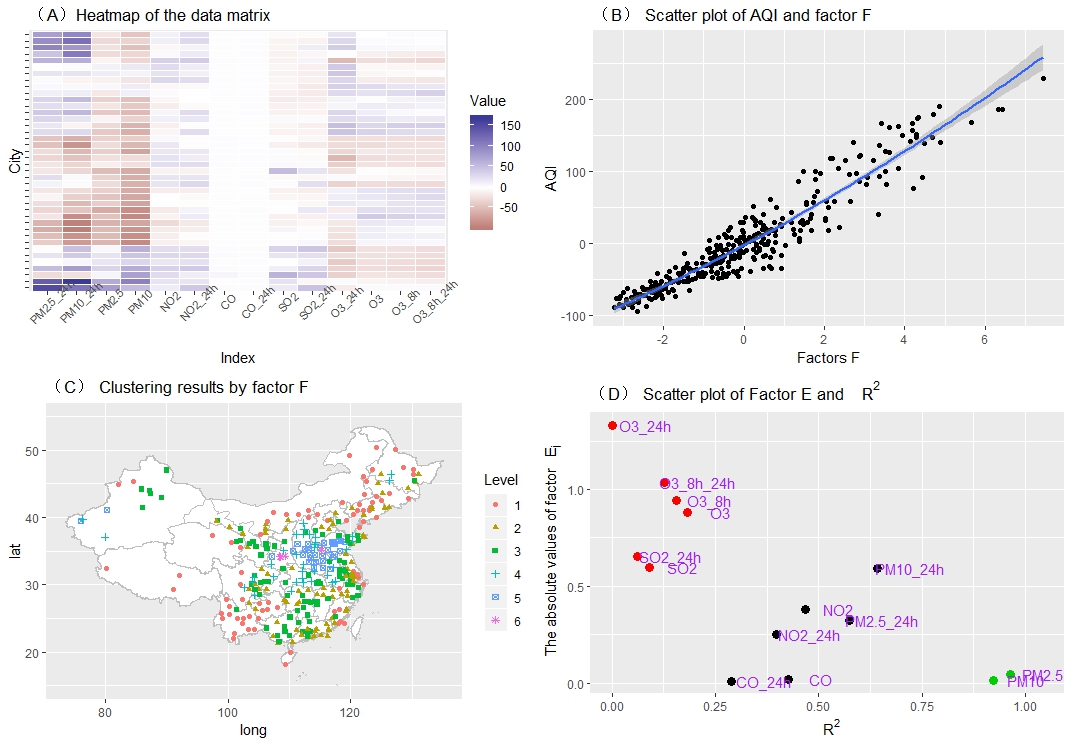}}\\
  \caption{\small (A) Heatmap for the air quality data, (B) Scatterplot of $\hat{F}_i$ vs AQIs for 338 cities in China,
  (C) K-means clustering results for $\hat{F}_i$, (D) Scatterplot of $\hat{E}_j$ vs $R^2_j$.}\label{R2}
  \normalsize
\end{figure}

\csection{DISCUSSION}

In this paper, we have developed a model-based method for
high-dimensional matrix data analysis. Our model, called 2wFM,
extends the application scope
of the classical factor model from traditional vector-valued data to a general kind of matrix-valued data,
in which there exists specific correlation structures between the
attributes of rows and those of columns. We construct the identification conditions for 2wFM,
derive an explicit expression to the likelihood function, and achieve
maximum likelihood estimation for each parameter. Under general conditions,
we study and obtain a series of theoretical studies
on the resulted estimators, including consistency properties as
well as asymptotic distributions.
Our results provide detailed discussions on the relationships between
the large sample behaviors of estimators and the statistical
properties of hidden factors.
Simulation studies further confirm our theoretical results and show
that our method can efficiently estimate parameters with high
precision.
Results on air quality data indicate that our method can extract and synthesize
information from both pollutants and geographical locations, and
can thus provide a more comprehensive evaluation of the level of air pollution
than the traditional AQI-based methods.
It would be of interest to develop a method that determines
the number of row and column factors based on some statistical
criteria such as BIC. Generalizations beyond the assumption of normal distributions for factors
and noises
may also be a considerable problem.

\scsection{SUPPLEMENTARY MATERIALS}
Supplementary material for ``A two-way factor model for
a high-dimensional data matrix''. The simulation results, rigorous proof of Theorem
1 and Theorem 2 in simple case, as well as
the relative lemmas with detailed proof that are useful in
proving Theorem 1-2 are all provided in the supplementary material.

\scsection{APPENDIX}

\textbf{\underline{Proof of Proposition 1}}.\\
It suffices to prove that if $\Sigma_{X} = \Sigma^{'}_{X}$ holds, then $\theta = \theta^{'}$.
Recall that $\Sigma_{X} = I_{p}\otimes A + B \otimes I_{q} + \sigma^{2}I_{p} \otimes I_{q}$, that is,
$$
\left(
\begin{array}{ccc}
  \Sigma_{11} & ... & \Sigma_{1p} \\
  \vdots & \ddots & \vdots \\
  \Sigma_{p1} & ... & \Sigma_{pp}
\end{array}
\right ) =  \left(
\begin{array}{cccc}
  A   &  &   \\
    &   \ddots  &  \\
    &      &   A
  \end{array}
\right )  + \left( \begin{array}{ccc}
               b_{11}I_{q}  & ... & b_{1p}I_{q} \\
                  \vdots& \ddots & \vdots \\
               b_{p1}I_{q} & ... & b_{pp}I_{q}
             \end{array} \right ) + \left(
\begin{array}{ccc}
  \sigma^2 I_{q}    &  &   \\
    &   \ddots  &  \\
    &      &   \sigma^2 I_{q}
  \end{array}
\right ),$$
where $\Sigma_{X} = (\Sigma_{ij})_{p \times p}$, $ \Sigma_{11},\dots,\Sigma_{pp}$ are $q \times q$ matrices, and $B=(b_{km})_{p \times p}$.
Since $\Sigma_{X} = \Sigma^{'}_{X}$, we have $\Sigma_{ii} = \Sigma^{'}_{ii}$, $i=1$, $\dots$, $p$, that is,
$$A + (b_{ii} + \sigma^2)I_{q} = A^{'} + (b^{'}_{ii} + \sigma^{'2})I_{q},i=1,\dots,p,$$
where $A = L\Psi_{F}L^{T}$, $A^{'} = L^{'}\Psi^{'}_{F}L^{'T}$ with
$\frac{L^{T}L}{q\sigma^{2}} = I_{r}$ and $\frac{L^{'T}L^{'}}{q\sigma^{'2}} = I_{r}$.
Note that $A + (b_{ii} + \sigma^2)I_{q}$ is  `a low-rank plus a diagonal matrix'
decomposition to $\Sigma_{ii}$. It is well known that when $\min\{p, q\} > \max\{r, c\}$,
this decomposition is unique under the identification conditions
(IC1) and (IC2) \citep{Anderson:(1956)}. This leads to:
\begin{eqnarray}\label{P1-1}
A = A^{'},~ b_{ii} + \sigma^2 = b^{'}_{ii} + \sigma^{'2}.
\end{eqnarray}
Because $\Sigma_{X} = \Sigma^{'}_{X}$, with $({\ref{P1-1}})$, we further have
$$B \otimes I_{q} + \sigma^{2} I_{p} \otimes I_{q} = B^{'} \otimes I_{q} + \sigma^{'2} I_{p} \otimes I_{q},$$
which implies $B + \sigma^{2} I_{p} = B^{'} + \sigma^{'2} I_{p}$. Note $B + \sigma^{2} I_{p}$ is also
`a low-rank plus a diagonal matrix' decomposition to a known matrix.
This leads to B = $B^{'}$, $\sigma^2 = \sigma^{'2}$.
With (IC1), (IC2) and the above results,
$\Lambda = \Lambda^{'}$, $\Psi_{E}= \Psi_{E}^{'}$, $L = L^{'}$ and $\Psi_{F}=\Psi_{F}^{'}$ are
immediate consequences of the identification conclusions of the classical factor model.
$\hfill\blacksquare$

\textbf{\underline{Proof of Proposition 2}}.\\
In simple case of $r=1$ and $c=1$, consider $\frac{\ln\ell(\theta;X)}{pq}$,
\begin{eqnarray*}
&&\dfrac{\ln\ell\left(L,\Lambda, \sigma^{2}_{F},\sigma^{2}_{E},\sigma^{2};X \right)}{pq}\nonumber \\
&&= -\dfrac{\ln(1+p\sigma^{2}_{E}+q\sigma^{2}_{F})}{pq} - \dfrac{(p-1)\ln(1+q\sigma^{2}_{F})}{pq}- \dfrac{(q-1)\ln(1+p\sigma^{2}_{E})}{pq} - \ln(\sigma^{2})\nonumber \\
&&~~~ - \dfrac{\tr(X^{T}X)}{pq\sigma^{2}} + \dfrac{\sigma^{2}_{F}L^{T}X^{T}XL}{pq\sigma^{4}(1+q\sigma^{2}_{F})} + \dfrac{\sigma^{2}_{E}\Lambda^{T}XX^{T}\Lambda}{pq\sigma^{4}(1+p\sigma^{2}_{E})} \nonumber \\
&&~~~ - \dfrac{L^{T}X^{T}\Lambda\Lambda^{T}XL}{p^2q^2\sigma^{6}}\left( 1- \dfrac{1}{1+q\sigma^{2}_{F}} - \dfrac{1}{1+p\sigma^{2}_{E}} +\dfrac{1}{1+q\sigma^{2}_{F}+p\sigma^{2}_{E}}  \right).
\end{eqnarray*}
When $\sigma^{2}\rightarrow\infty$, $-\ln\sigma^{2}\rightarrow -\infty$. On the other hand,
the remaining items in  $\frac{\ln\ell(\theta;X)}{pq}$ are bounded in probability. Thus, there exists a large enough constant $C_{0}$ such that $\hat{\sigma}^{2}\in \left\{\sigma^{2}:\sigma^{2}\leq C_{0} \right\}$ in probability.
Next, it is shown that there exists a lower bound $C_1$, such that
$\hat{\sigma}^{2}\in \left\{\sigma^{2}:\sigma^{2}\geq C_{1} \right\}$ in probability.
Let $\tilde{L}=\frac{L}{\sqrt{q\sigma^{2}}}$, $\tilde{\Lambda}=\frac{\Lambda}{\sqrt{p\sigma^{2}}}$, we have
\begin{eqnarray}\label{P2-2}
&&\tr[\Sigma^{-1}_{X}\vec(X)\vec^{T}(X)]\nonumber \\
&&=\dfrac{1}{\sigma^{2}}\tr\left(X^{T}X\right) - \dfrac{\sigma^{2}_{F}}{\sigma^{4}(1+q\sigma^{2}_{F})}L^{T}X^{T}XL - \dfrac{\sigma^{2}_{E}}{\sigma^{4}(1+p\sigma^{2}_{E})}\Lambda^{T}XX^{T}\Lambda\nonumber \\
&&~~+ \dfrac{L^{T}X^{T}\Lambda\Lambda^{T}XL}{pq\sigma^{6}}\left( 1- \dfrac{1}{1+q\sigma^{2}_{F}}-\dfrac{1}{1+p\sigma^{2}_{E}} + \dfrac{1}{1+q\sigma^{2}_{F}+p\sigma^{2}_{E}}\right)\nonumber \\
&&=\dfrac{1}{\sigma^{2}}\left[\tr\left(X^{T}X\right) - \tilde{L}^{T}X^{T}X\tilde{L} - \tilde{\Lambda}^{T}XX^{T}\tilde{\Lambda} + \tilde{L}^{T}X^{T}\tilde{\Lambda}\tilde{\Lambda}^{T}X\tilde{L} \right]\nonumber \\
&&~~ + \dfrac{1}{\sigma^{2}}\left[ \dfrac{\tilde{L}^{T}X^{T}X\tilde{L}-\tilde{L}^{T}X^{T}\tilde{\Lambda}\tilde{\Lambda}^{T}X\tilde{L}}{1+q\sigma^{2}_{F}} + \dfrac{\tilde{\Lambda}^{T}XX^{T}\tilde{\Lambda}-\tilde{\Lambda}^{T}X\tilde{L}\tilde{L}^{T}X^{T}\tilde{\Lambda}}{1+p\sigma^{2}_{E}}\right.\nonumber \\
&&~~~~\left. + \dfrac{\tilde{\Lambda}^{T}X\tilde{L}\tilde{L}^{T}X^{T}\tilde{\Lambda}}{1+p\sigma^{2}_{E}+q\sigma^{2}_{F}} \right]\nonumber \\
&&\triangleq g^{tr}_{1}(\theta) + g^{tr}_{2}(\theta),
\end{eqnarray}
where
\begin{eqnarray*}
g^{tr}_{1}(\theta)&=&\dfrac{1}{\sigma^{2}}\left[\tr\left(X^{T}X\right) - \tilde{L}^{T}X^{T}X\tilde{L} - \tilde{\Lambda}^{T}XX^{T}\tilde{\Lambda} + \tilde{L}^{T}X^{T}\tilde{\Lambda}\tilde{\Lambda}^{T}X\tilde{L} \right],\nonumber \\
g^{tr}_{2}(\theta) &=& \dfrac{1}{\sigma^{2}}\left[ \dfrac{\tilde{L}^{T}X^{T}X\tilde{L}-\tilde{L}^{T}X^{T}\tilde{\Lambda}\tilde{\Lambda}^{T}X\tilde{L}}{1+q\sigma^{2}_{F}} + \dfrac{\tilde{\Lambda}^{T}XX^{T}\tilde{\Lambda}-\tilde{\Lambda}^{T}X\tilde{L}\tilde{L}^{T}X^{T}\tilde{\Lambda}}{1+p\sigma^{2}_{E}}\right. \nonumber \\
&&~~~~\left. + \dfrac{\tilde{\Lambda}^{T}X\tilde{L}\tilde{L}^{T}X^{T}\tilde{\Lambda}}{1+p\sigma^{2}_{E}+q\sigma^{2}_{F}} \right].
\end{eqnarray*}
For $g^{tr}_{1}$, by Lemma 6A and the expression of $\tr(X^{T}X)$ in Lemma 12, it can be verified that
\begin{eqnarray}\label{P2-3}
\dfrac{g^{tr}_{1}(\theta)}{pq}&=&\dfrac{1}{pq\sigma^{2}}\left[\tr\left(X^{T}X\right) - \tilde{L}^{T}X^{T}X\tilde{L} - \tilde{\Lambda}^{T}XX^{T}\tilde{\Lambda} + \tilde{L}^{T}X^{T}\tilde{\Lambda}\tilde{\Lambda}^{T}X\tilde{L} \right]\nonumber \\
&\geqslant& \dfrac{\sigma^{*2}\left[1+o_{p}(1)\right]}{\sigma^{2}},
\end{eqnarray}
For $g^{tr}_{2}$, we have
\begin{eqnarray*}\label{P2-4}
&\tilde{L}^{T}X^{T}X\tilde{L}-\tilde{L}^{T}X^{T}\tilde{\Lambda}\tilde{\Lambda}^{T}X\tilde{L} = \tilde{L}^{T}X^{T}\left(I_{p}-\tilde{\Lambda}\tilde{\Lambda}^{T}\right)X\tilde{L} \geqslant 0, \\ &\tilde{\Lambda}^{T}XX^{T}\tilde{\Lambda}-\tilde{\Lambda}^{T}X\tilde{L}\tilde{L}^{T}X^{T}\tilde{\Lambda}= \tilde{\Lambda}^{T}X\left(I_{q}-\tilde{L}\tilde{L}^{T}\right)X^{T}\tilde{\Lambda} \geqslant 0, \\
&\tilde{L}^{T}X^{T}\tilde{\Lambda}\tilde{\Lambda}^{T}X\tilde{L}\geqslant 0.
\end{eqnarray*}
Moreover, as $\sigma^{2}\rightarrow0^{+}$,
\begin{eqnarray}\label{P2-5}
-\dfrac{g^{tr}_{2}(\theta)}{pq} \leq 0.
\end{eqnarray}
With (\ref{P2-2}) and (\ref{P2-5}),
as $p\rightarrow\infty$ and $q\rightarrow\infty$, only concerned on $\sigma^{2}$, the log-likelihood function can be expressed as
\begin{eqnarray*}\label{P2-8}
\dfrac{\ln\ell(\theta)}{pq} &=& -\dfrac{\ln|\Sigma|}{pq} - \dfrac{tr[\Sigma^{-1}vec(X)vec^{T}(X)]}{pq}\nonumber \\
&=& -\dfrac{\ln\left(1+q\sigma^{2}_{F}+p\sigma^{2}_{E}\right)}{pq} - \dfrac{(p-1)\ln(1+q\sigma^{2}_{F})}{pq} - \dfrac{(q-1)\ln(1+p\sigma^{2}_{E})}{pq}\nonumber \\
&&- \ln\sigma^{2} - \dfrac{ g^{tr}_{1}(\theta) + g^{tr}_{2}(\theta)}{pq}\nonumber \\
&\propto& - \ln\sigma^{2} - \dfrac{g^{tr}_{1}(\theta)}{pq} - \dfrac{g^{tr}_{2}(\theta)}{pq}.
\end{eqnarray*}
With $(\ref{P2-3})$, as $\sigma^{2}\rightarrow0^{+}$, there exists a small enough probability $\delta_{pq}$ satisfying $\lim_{p,q\rightarrow \infty}\delta_{pq}=0^{+}$, such that
\begin{eqnarray}\label{P2-9}
P\left(- \ln\sigma^{2} - \dfrac{g^{tr}_{1}(\theta)}{pq} \leq - \ln\sigma^{2} - \dfrac{\sigma^{*2}\left[1+o_{p}(1)\right]}{\sigma^{2}}\rightarrow -\infty\right)= 1-\delta_{pq}.
\end{eqnarray}
Summarizing $(\ref{P2-5})$ and $(\ref{P2-9})$, as $\sigma^{2}\rightarrow0^{+}$, we have
\begin{eqnarray*}
P\left(\dfrac{\ln\ell(\theta)}{pq}\rightarrow -\infty\right)= 1-\delta_{pq},
\end{eqnarray*}
which means that as $p\rightarrow\infty$ and $q\rightarrow\infty$, there exists a small enough constant $C_{1}$ such that
$\hat{\sigma}^{2}\in \left\{\sigma^{2}:\sigma^{2}\geq C_{1} \right\}$ in probability.
Moreover, Lemma 7 shows that there exists a large enough constant $C_{2}$, such that $\left( \hat{\sigma}^{2}_{F},\hat{\sigma}^{2}_{E} \right)^{T} \in \left[C^{-1}_{2},C_{2}\right]^{2}$ in probability. We therefore conclude that
we can let $\tilde{C}=\max\{C_{0}, C^{-1}_{1},C_{2}\}$  such that
 \begin{eqnarray*}
\left( \hat{\sigma}^{2}_{F},\hat{\sigma}^{2}_{E},\hat{\sigma}^{2} \right)^{T} \in \left[\tilde{C}^{-1},\tilde{C}\right]^{3}~ \text{in probability}.
\end{eqnarray*}

For general case, that is, $r \geq 1$, $c \geq 1$, the conclusion
$$\left( \diag^{T}\left(\hat{\Psi}_{F}\right),\diag^{T}\left(\hat{\Psi}_{E}\right),\hat{\sigma}^{2} \right)^{T} \in \left[\tilde{C}^{-1},\tilde{C}\right]^{r+c+1}~ \text{in probability}$$
can be obtained in a similar way.
$\hfill\blacksquare$\\

\textbf{\underline{Proof of Theorem 1}}.\\
With the definition of MLE, we have,
for $\left( \hat{L},\hat{\Lambda} \right)$,
\begin{eqnarray}\label{T3.2}
\left(\hat{L},\hat{\Lambda}\right) = \mathop{\arg\max}_{L^{T}L=q\hat{\sigma}^{2}I_{r},\Lambda^{T}\Lambda=p\hat{\sigma}^{2}I_{c}}{\ln\ell\left( L,\Lambda,\hat{\Psi}_{F},\hat{\Psi}_{E},\hat{\sigma}^{2} ;X \right)},
\end{eqnarray}
and for $\left( \hat{\Psi}_{F},\hat{\Psi}_{E},\hat{L},\hat{\Lambda} \right)$,
\begin{eqnarray}\label{T3.1}
\left(\hat{\Psi}_{F},\hat{\Psi}_{E},\hat{L},\hat{\Lambda} \right) = \mathop{\arg\max}_{\left( L,\Lambda \right)~s.t.~(\ref{T3.2})} {\ln\ell\left( L,\Lambda,\Psi_{F},\Psi_{E},\hat{\sigma}^{2} ;X \right)}.
\end{eqnarray}
With (\ref{T3.2}) and Proposition 4, from Lemma 7, we have
\begin{eqnarray}\label{T3.11}
&&\mathop{\max}_{L^{T}L=q\hat{\sigma}^{2}I_{r},\Lambda^{T}\Lambda=p\hat{\sigma}^{2}I_{c}}{\ln\ell\left( L,\Lambda,\hat{\Psi}_{F},\hat{\Psi}_{E},\hat{\sigma}^{2} ;X \right)}\nonumber\\
&&=\mathop{\max}_{(IC1),(IC2)}\left( \dfrac{ \sum_{j=1}^{r}\hat{\sigma}^{2}_{F_{j}}\hat{d}_{2j}L_{j}^{T}X^{T}XL_{j}}{pq} + \dfrac{\sum_{i=1}^{c}\hat{\sigma}^{2}_{E_{i}}\hat{d}_{3i}\Lambda_{i}^{T}XX^{T}\Lambda_{i}}{pq}\right.\nonumber \\
&&\left.~~~~~~~~~~~~~~~~~~~~~~ - \dfrac{\sum_{i=1}^{c}\sum_{j=1}^{r}\hat{\sigma}^{2}_{F_{j}}\hat{\sigma}^{2}_{E_{i}}\hat{d}_{4ij}L_{j}^{T}X^{T}\Lambda_{i}\Lambda_{i}^{T}XL_{j}}{pq} \right)\nonumber \\
&&~~~~ = \dfrac{\sigma^{*2}}{\hat{\sigma}^{2}}\left( \sum_{j=1}^{r}\sigma^{*2}_{F_{j}} + \sum_{i=1}^{c}\sigma^{*2}_{E_{i}} \right) + o_{p}(1),
\end{eqnarray}
The optimization function on $\left(\hat{\Psi}_{F},\hat{\Psi}_{E},\hat{L},\hat{\Lambda}\right)$ in $(\ref{T3.1})$ is the same within Lemma 7. Thus, we get
\begin{eqnarray}\label{T3.5}
\dfrac{\hat{L}^{T}X^{T}\hat{\Lambda}}{pq\hat{\sigma}^{4}}= o_{p}(1)\textbf{1}_{r\times c},
\end{eqnarray}
\begin{eqnarray}\label{T3.3}
\diag\left( \hat{\Psi}_{F} \right) = \diag\left( \dfrac{\hat{L}^{T}X^{T}X\hat{L}}{pq^{2}\hat{\sigma}^{4}} \right) + o_{p}(1)\textbf{1}_{r},
\end{eqnarray}
and
\begin{eqnarray}\label{T3.4}
\diag\left(\hat{\Psi}_{E}\right) = \diag\left( \dfrac{\hat{\Lambda}^{T}XX^{T}\hat{\Lambda}}{p^{2}q\hat{\sigma}^{4}}\right) + o_{p}(1)\textbf{1}_{c}.
\end{eqnarray}
Therefore, in (\ref{T3.3}), the diagonal elements have been proven and the non-diagonal elements also meet the equation, as shown in what follows.
The estimating equation of $L_{j}$ shown in Subsection A.8 in supplement material, pre-multiplies $\frac{\hat{\sigma}^{2}_{F_{j_{1}}}\hat{L}_{j_{1}}^{T}}{p}(j_{1}\neq j)$, giving us
\begin{eqnarray}\label{T3.6}
0 &=& \hat{\sigma}^{2}_{F_{j_{1}}}\hat{w}^{j}_{2j_{1}}\dfrac{\hat{L}_{j_{1}}^{T}\hat{L}_{j_{1}}}{p} - \sum_{k=1}^{c}\hat{\sigma}^{2}_{F_{j_{1}}}\hat{w}_{3j_{1}k}^{j}\dfrac{\hat{L}_{j_{1}}^{T}\hat{L}_{j_{1}}}{p} - \hat{\sigma}^{2}_{F_{j_{1}}}\hat{\Delta}_{1}^{j}\dfrac{\hat{L}_{j_{1}}^{T}X^{T}X\hat{L}_{j}}{p}\nonumber \\
&&+ \sum_{k=1}^{c}\hat{\sigma}^{2}_{F_{j_{1}}}\hat{\Delta}_{2k}^{j} \dfrac{\hat{L}_{j_{1}}^{T}X^{T}\hat{\Lambda}_{k}\hat{\Lambda}_{k}^{T}X\hat{L}_{j}}{p}\nonumber \\
&=&\dfrac{q\hat{\sigma}^{2}_{F_{j}}\hat{\sigma}^{4}_{F_{j_{1}}}\hat{L}^{T}_{j}X^{T}X\hat{L}_{j_{1}}} {p\hat{\sigma}^{4}(1+q\hat{\sigma}^{2}_{F_{j}})(1+q\hat{\sigma}^{2}_{F_{j_{1}}})}
- \dfrac{\hat{\sigma}^{2}_{F_{j}}\hat{\sigma}^{2}_{F_{j_{1}}}\hat{L}_{j_{1}}^{T}X^{T}X\hat{L}_{j}} {p\hat{\sigma}^{4}(1+q\hat{\sigma}^{2}_{F_{j}})}\nonumber \\
&& - \sum_{k=1}^{c}\left[\hat{\sigma}^{2}_{F_{j_{1}}}\hat{w}_{3j_{1}k}^{j}\dfrac{\hat{L}_{j_{1}}^{T}\hat{L}_{j_{1}}}{p}- \hat{\sigma}^{2}_{F_{j_{1}}}\hat{\Delta}_{2k}^{j} \dfrac{\hat{L}_{j_{1}}^{T}X^{T}\hat{\Lambda}_{k}\hat{\Lambda}_{k}^{T}X\hat{L}_{j}}{p}\right]\nonumber \\
&=& - \dfrac{\hat{\sigma}^{2}_{F_{j}}\hat{\sigma}^{2}_{F_{j_{1}}}\hat{L}^{T}_{j}X^{T}X\hat{L}_{j_{1}}} {p\hat{\sigma}^{4}(1+q\hat{\sigma}^{2}_{F_{j}})(1+q\hat{\sigma}^{2}_{F_{j_{1}}})}+ \sum_{k=1}^{c} \dfrac{\hat{L}^{T}_{j}X^{T}\hat{\Lambda}_{k}\hat{\Lambda}_{k}^{T}X\hat{L}_{j_{1}}} {p^{2}\hat{\sigma}^{6}}\left[ \dfrac{\hat{\sigma}^{2}_{F_{j_{1}}}\hat{\sigma}^{2}_{F_{j}}}{(1+q\hat{\sigma}^{2}_{F_{j}})(1+q\hat{\sigma}^{2}_{F_{j_{1}}})}\right.\nonumber\\
&&\left.~~~~ - \dfrac{\hat{\sigma}^{2}_{F_{j_{1}}}\hat{\sigma}^{2}_{F_{j}}} {(1+q\hat{\sigma}^{2}_{F_{j}}+p\hat{\sigma}^{2}_{E_{k}})(1+q\hat{\sigma}^{2}_{F_{j_{1}}}+p\hat{\sigma}^{2}_{E_{k}})} \right],
\end{eqnarray}
where $\hat{w}_{1}^{j}$, $ \hat{w}^{j}_{2s} $, $ \hat{w}_{3sk}^{j} $, $ \hat{\Delta}_{1}^{j} $ and $\hat{\Delta}_{2k}^{j}$ are all scalars
with the detailed expressions shown in Subsection A.8.
Through Proposition 4 and $(\ref{T3.5})$, (\ref{T3.6}) could be modified as
\begin{eqnarray*}\label{T3.7}
&&0= - \dfrac{\hat{L}^{T}_{j}X^{T}X\hat{L}_{j_{1}}} {pq^{2}\hat{\sigma}^{4}}\left[ 1+ o_{p}(1) \right] + \sum_{k=1}^{c} \dfrac{\hat{L}^{T}_{j}X^{T}\hat{\Lambda}_{k}\hat{\Lambda}_{k}^{T}X\hat{L}_{j_{1}}} {p^{2}q^{2}\hat{\sigma}^{6}}\bigg[ 1 + o_{p}(1) \bigg.\nonumber \\
&&\left. ~~~~~~~~ - \dfrac{\hat{\sigma}^{2}_{F_{j_{1}}}\hat{\sigma}^{2}_{F_{j}}} {(q\hat{\sigma}^{2}_{F_{j}}+p\hat{\sigma}^{2}_{E_{k}})(q\hat{\sigma}^{2}_{F_{j_{1}}}+p\hat{\sigma}^{2}_{E_{k}})}\right]\nonumber \\
&&~~= \dfrac{\hat{L}^{T}_{j}X^{T}X\hat{L}_{j_{1}}} {pq^{2}\hat{\sigma}^{4}}\left[ 1+ o_{p}(1) \right] + o_{p}(1),\nonumber \\
&& \Rightarrow \dfrac{\hat{L}^{T}_{j}X^{T}X\hat{L}_{j_{1}}} {pq^{2}\hat{\sigma}^{4}} = o_{p}(1),~j_{1}\neq j.
\end{eqnarray*}
Then, with $(\ref{T3.3})$, we have
\begin{eqnarray}\label{T3.8}
\hat{\Psi}_{F} = \dfrac{\hat{L}^{T}X^{T}X\hat{L}}{pq^{2}\hat{\sigma}^{4}} + o_{p}(1)\textbf{1}_{r\times r}.
\end{eqnarray}
Similarly, we have
\begin{eqnarray}\label{T3.9}
\hat{\Psi}_{E} =  \dfrac{\hat{\Lambda}^{T}XX^{T}\hat{\Lambda}}{p^{2}q\hat{\sigma}^{4}} + o_{p}(1)\textbf{1}_{c\times c}.
\end{eqnarray}
Through $(\ref{T3.5})$, $(\ref{T3.8})$, and $(\ref{T3.9})$, $\left(\hat{L},\hat{\Lambda}\right)$
now satisfies the conditions in Lemma 8, that is,
\begin{eqnarray}\label{T3.91}
\left( \begin{array}{c} \hat{\alpha}\\ \hat{\beta} \\ \end{array} \right) = \left( \begin{array}{cc} \hat{\alpha}_{L^{*}}~~\hat{\alpha}_{E}\\ \hat{\beta}_{F} ~~~ \hat{\beta}_{\Lambda^{*}} \\ \end{array}\right) = \left( \begin{array}{cc} \frac{L^{*T}\hat{L}}{q\sigma^{*}\hat{\sigma}}~~\frac{\tilde{E}^{T}\hat{L}} {q\sigma^{*}\hat{\sigma}}\\ \\ \frac{\tilde{F}^{T}\hat{\Lambda}}{p\sigma^{*}\hat{\sigma}} ~~ \frac{\Lambda^{*T}\hat{\Lambda}}{p\sigma^{*}\hat{\sigma}} \\ \end{array}\right)  = I_{r+c} + o_{p}(1)\textbf{1}_{(r+c) \times (r+c)},
\end{eqnarray}
where
$\hat{\alpha}_{L^{*}} = \frac{L^{*T}\hat{L}}{q\sigma^{*}\hat{\sigma}},~
\hat{\alpha}_{E} = \frac{\tilde{E}^{T}\hat{L}}{q\sigma^{*}\hat{\sigma}},~
\hat{\beta}_{F} = \frac{\tilde{F}^{T}\hat{\Lambda}}{p\sigma^{*}\hat{\sigma}},~
\hat{\beta}_{\Lambda^{*}} = \frac{\Lambda^{*T}\hat{\Lambda}}{p\sigma^{*}\hat{\sigma}},
\hat{\alpha} = \left( \hat{\alpha}_{L^{*}}~ \hat{\alpha}_{E} \right),~\hat{\beta} = \left( \hat{\beta}_{F}~ \hat{\beta}_{\Lambda^{*}} \right)$.
In the above, we have proven $\frac{L^{*T}\hat{L}}{q\sigma^{*}\hat{\sigma}}=I_{r}+o_{p}(1)\textbf{1}_{r\times r}$ and $\frac{\Lambda^{*T}\hat{\Lambda}}{p\sigma^{*}\hat{\sigma}}=I_{c}+o_{p}(1)\textbf{1}_{c\times c}$. In what follows, we will show the consistency of $\hat{\sigma}^{2}$.

Recall the estimating equation of $\sigma^2$ in Subsection A.8 and by multiplying $\frac{\hat{\sigma}^{2}}{pq}$, the equation could be written as
\begin{eqnarray}\label{T3.14}
&&-\sum_{i=1}^{c}\sum_{j=1}^{r}\dfrac{1}{pq( 1+ q\hat{\sigma}^{2}_{F_{j}} + p\hat{\sigma}^{2}_{E_{i}})} - \sum_{j=1}^{r}\dfrac{p-c}{pq(1+q\hat{\sigma}^{2}_{F_{j}})} - \sum_{i=1}^{c}\dfrac{q-r}{pq(1+p\hat{\sigma}^{2}_{E_{i}})} \nonumber \\
&&- \dfrac{(p-c)(q-r)}{pq} + \dfrac{\tr\left(X^{T}X\right)}{pq\hat{\sigma}^{2}} - \sum_{j=1}^{r}\left[1 - \dfrac{1}{(1+q\hat{\sigma}^{2}_{F_{j}})^2} \right]\dfrac{\hat{L}_{j}^{T}X^{T}X\hat{L}_{j}}{pq^2\hat{\sigma}^{4}}\nonumber \\
&&- \sum_{i=1}^{c}\left[1 - \dfrac{1}{(1+p\hat{\sigma}^{2}_{E_{i}})^2}
\right]\dfrac{\hat{\Lambda}_{i}^{T}XX^{T}\hat{\Lambda}_{i}}{p^2q\hat{\sigma}^{4}} + \sum_{i=1}^{c}\sum_{j=1}^{r}\left[1 - \dfrac{1}{(1+q\hat{\sigma}^{2}_{F_{j}})^2} - \dfrac{1}{(1+p\hat{\sigma}^{2}_{E_{i}})^2}\right.\nonumber \\
&&~~~~ \left. + \dfrac{1}{(1+q\hat{\sigma}^{2}_{F_{j}}+ p\hat{\sigma}^{2}_{E_{i}})^2} \right] \dfrac{\hat{L}_{j}^{T}X^{T}\hat{\Lambda}_{i} \hat{\Lambda}_{i}^{T}X \hat{L}_{j}}{p^2q^2\hat{\sigma}^{6}} = 0.
\end{eqnarray}
Using Proposition 4 and the techniques used during the proof of Lemma 5 and Lemma 6B, as $p,q\rightarrow \infty$, it can be verified that
\begin{eqnarray*}
&&-\sum_{i=1}^{c}\sum_{j=1}^{r}\dfrac{1}{pq( 1+ q\hat{\sigma}^{2}_{F_{j}} + p\hat{\sigma}^{2}_{E_{i}})} - \sum_{j=1}^{r}\dfrac{p-1}{pq(1+q\hat{\sigma}^{2}_{F_{j}})} - \sum_{i=1}^{c}\dfrac{q-1}{pq(1+p\hat{\sigma}^{2}_{E_{i}})} =o_{p}(1),\\
&&\sum_{j=1}^{r}\dfrac{1}{(1+q\hat{\sigma}^{2}_{F_{j}})^2} \dfrac{\hat{L}_{j}^{T}X^{T}X\hat{L}_{j}}{pq^2\hat{\sigma}^{4}} = O_{p}(p^{-2}) = o_{p}(1),\nonumber \\
&&\sum_{i=1}^{c}\dfrac{1}{(1+p\hat{\sigma}^{2}_{E_{i}})^2} \dfrac{\hat{\Lambda}_{i}^{T}XX^{T}\hat{\Lambda}_{i}}{p^2q\hat{\sigma}^{4}} = O_{p}(p^{-2}) = o_{p}(1),\nonumber \\
&&\sum_{i=1}^{c}\sum_{j=1}^{r}\left[-\dfrac{1}{(1+q\hat{\sigma}^{2}_{F_{j}})^2} - \dfrac{1}{(1+p\hat{\sigma}^{2}_{E_{i}})^2}\right.\\
&&~~~~~~~~~~~~\left.+ \dfrac{1}{(1+q\hat{\sigma}^{2}_{F_{j}} + p\hat{\sigma}^{2}_{E_{i}})^2} \right] \dfrac{\hat{L}_{j}^{T}X^{T}\hat{\Lambda}_{i} \hat{\Lambda}_{i}^{T}X \hat{L}_{j}}{p^2q^2\hat{\sigma}^{6}} =o_{p}(1).
\end{eqnarray*}
Putting the above conclusions into $(\ref{T3.14})$, we obtain
\begin{eqnarray}\label{T3.19}
&&-1 + \dfrac{\tr\left(X^{T}X\right)}{pq\hat{\sigma}^{2}} -\sum_{j=1}^{r}\dfrac{\hat{L}_{j}^{T}X^{T}X\hat{L}_{j}}{pq^2\hat{\sigma}^{4}} - \sum_{i=1}^{c}\dfrac{\hat{\Lambda}_{i}^{T}XX^{T}\hat{\Lambda}_{i}}{p^2q\hat{\sigma}^{4}}\nonumber \\
&&+ \sum_{i=1}^{c}\sum_{j=1}^{r}\dfrac{\hat{L}_{j}^{T}X^{T}\hat{\Lambda}_{i} \hat{\Lambda}_{i}^{T}X \hat{L}_{j}}{p^2q^2\hat{\sigma}^{6}} =o_{p}(1).
\end{eqnarray}
With $(\ref{T3.11})$, we further have
\begin{equation}\label{T3.20}
-1 + \dfrac{\tr\left(X^{T}X\right)}{pq\hat{\sigma}^{2}} -\dfrac{\sigma^{*2}}{\hat{\sigma}^{2}}\left(\sum_{j=1}^{r}\sigma^{*2}_{F_{j}}+ \sum_{i=1}^{c}\sigma^{*2}_{E_{i}} \right) = o_{p}(1).
\end{equation}
From Lemma 4, we have
\begin{eqnarray}\label{T3.21}
\dfrac{\tr\left(X^{T}X\right)}{pq\hat{\sigma}^2} &=& \dfrac{1}{pq\hat{\sigma}^2}\tr\left[\left( FL^{*T} + \Lambda^{*}E^{T} + \epsilon \right)^{T} \left( FL^{*T} + \Lambda^{*}E^{T} + \epsilon \right) \right]\nonumber \\
&=& \dfrac{\tr\left( F^{T}FL^{*T}L^{*}\right)}{pq\hat{\sigma}^2} + \dfrac{\tr\left( E^{T}E\Lambda^{*T}\Lambda^{*}\right)}{pq\hat{\sigma}^2} +\dfrac{\tr(\epsilon^{T}\epsilon)}{pq\hat{\sigma}^2} + 2\dfrac{\tr\left( F^{T}\Lambda^{*}E^{T}L^{*}\right)}{pq\hat{\sigma}^2}\nonumber \\
&&+ 2\dfrac{\tr\left(F^{T}\epsilon L^{*} \right)}{pq\hat{\sigma}^2} + 2\dfrac{\tr\left(\Lambda^{*T}\epsilon E \right)}{pq\hat{\sigma}^2}  \nonumber \\
&=& \dfrac{\sigma^{*2}}{\hat{\sigma}^{2}}\left( \sum_{j=1}^{r}\sigma^{*2}_{F_{j}} + \sum_{i=1}^{c}\sigma^{*2}_{E_{i}} + 1 \right) + o_{p}(1).
\end{eqnarray}
Putting $(\ref{T3.21})$ into $(\ref{T3.20})$, with the bounded property of $\hat{\sigma}^{2}$, we obtain
\begin{eqnarray*}\label{T3.22}
o_{p}(1) &=& -1 + \dfrac{tr\left(X^{T}X\right)}{pq\hat{\sigma}^{2}} -\dfrac{\sigma^{*2}}{\hat{\sigma}^{2}}\left(\sum_{j=1}^{r}\sigma^{*2}_{F_{j}}+ \sum_{i=1}^{c}\sigma^{*2}_{E_{i}} \right)\nonumber \\
&=&   -1 +\dfrac{\sigma^{*2}}{\hat{\sigma}^{2}} \left( \sum_{j=1}^{r}\sigma^{*2}_{F_{j}}+ \sum_{i=1}^{c}\sigma^{*2}_{E_{i}} + 1 \right)  -\dfrac{\sigma^{*2}}{\hat{\sigma}^{2}}\left( \sum_{j=1}^{r}\sigma^{*2}_{F_{j}}+ \sum_{i=1}^{c}\sigma^{*2}_{E_{i}} \right)  + o_{p}(1) \nonumber \\
&=& -1 + \dfrac{\sigma^{*2}}{\hat{\sigma}^{2}} + o_{p}(1) = - \dfrac{\hat{\sigma}^2 - \sigma^{*2}}{\hat{\sigma}^{2}} + o_{p}(1),
\end{eqnarray*}
then
\begin{eqnarray}\label{T3.23}
\hat{\sigma}^{2}- \sigma^{*2}=o_{p}(1).
\end{eqnarray}
With (\ref{T3.23}), $(\ref{T3.91})$ says that
\begin{eqnarray*}\label{T3.25}
\dfrac{L^{*T}\hat{L}}{q\hat{\sigma}^{2}} = I_{r}+ o_{p}(1)\textbf{1}_{r\times r},~\dfrac{\Lambda^{*T}\hat{\Lambda}}{p\hat{\sigma}^{2}} = I_{c}+ o_{p}(1)\textbf{1}_{c\times c},~\dfrac{\tilde{E}^{T}\hat{L}} {q\hat{\sigma}^{2}} = o_{p}(1)\textbf{1}_{c\times r},~\dfrac{\tilde{F}^{T}\hat{\Lambda}}{p\hat{\sigma}^{2}} = o_{p}(1)\textbf{1}_{r\times c}.
\end{eqnarray*}
Thus,
\begin{eqnarray*}\label{T3.27}
\tr\left[ \dfrac{(\hat{L}-L^{*})^{T}(\hat{L}-L^{*})}{q\hat{\sigma}^{2}} \right] =\tr\left[I_{r} + \dfrac{\sigma^{*2}}{\hat{\sigma}^{2}}I_{r} - \dfrac{L^{*T}\hat{L}}{q\hat{\sigma}^{2}} - \dfrac{\hat{L}^{T}L^{*}}{q\hat{\sigma}^{2}}\right] = o_{p}(1),\\
\tr\left[\dfrac{(\hat{\Lambda}-\Lambda^{*})^{T}(\hat{\Lambda}-\Lambda^{*})}{p\hat{\sigma}^{2}}\right] = \tr\left[I_{c} + \dfrac{\sigma^{*2}}{\hat{\sigma}^{2}}I_{c} - \dfrac{\Lambda^{*T}\hat{\Lambda}}{p\hat{\sigma}^{2}} - \dfrac{\hat{\Lambda}^{T}\Lambda^{*}}{p\hat{\sigma}^{2}} \right] = o_{p}(1).
\end{eqnarray*}
With $(\ref{T3.8})$ and $(\ref{T3.9})$, consistent with $\hat{\sigma}^2$, $\hat{L}$ and $\hat{\Lambda}$,  from Lemma 9, we have
\begin{eqnarray*}\label{T3.29}
&&\hat{\Psi}_{F} = \dfrac{\hat{L}^{T}X^{T}X\hat{L}}{pq^{2}\hat{\sigma}^{4}} + o_{p}(1)\textbf{1}_{r\times r}=\Psi^{*}_{F} + o_{p}(1)\textbf{1}_{r\times r} \Rightarrow \hat{\Psi}_{F} - \Psi^{*}_{F} = o_{p}(1)\textbf{1}_{r\times r},\\
&&\hat{\Psi}_{E} = \dfrac{\hat{\Lambda}^{T}XX^{T}\hat{\Lambda}}{p^{2}q\hat{\sigma}^{4}}+o_{p}(1)\textbf{1}_{c\times c}=\Psi^{*}_{E} + o_{p}(1)\textbf{1}_{c\times c}\Rightarrow \hat{\Psi}_{E} - \Psi^{*}_{E} = o_{p}(1)\textbf{1}_{c\times c}.
\end{eqnarray*}
In summary, we get the consistency conclusions for $\hat{\theta}$, that is,
\begin{eqnarray*}\label{T3.31}
\hat{\Psi}_{F} - \Psi^{*}_{F} \stackrel{p}{\longrightarrow}  \textbf{0}_{r\times r} ,~
\hat{\Psi}_{E} - \Psi^{*}_{E} \stackrel{p}{\longrightarrow}  \textbf{0}_{c\times c} ,~
\hat{\sigma}^{2} - \sigma^{*2} \stackrel{p}{\longrightarrow}  0,\\
tr\left[\dfrac{(\hat{L}-L^{*})^{T}(\hat{L}-L^{*})}{q\hat{\sigma}^2}\right] \stackrel{p}{\longrightarrow} 0 ,~
tr\left[\dfrac{(\hat{\Lambda}-\Lambda^{*})^{T}(\hat{\Lambda}-\Lambda^{*})}{p\hat{\sigma}^2}\right] \stackrel{p}{\longrightarrow} 0.
\end{eqnarray*}

For the case for which there
exists one pair $(i_0,j_0)$, $i_0\in\{1,\dots,c\}$, $j_0\in \{1,\dots,r\}$,
such that $p^{2}\sigma^{2}_{E_{i_{0}}} = q^{2}\sigma^{2}_{F_{j_{0}}}$, or $M$ pairs ($M\leq \min\{r,c\}$),
$(i_{1},j_{1}), \dots,(i_{M},j_{M})$, $1\leq i_{1}<\cdots<i_{M}\leq c$, $1\leq j_{1}<\cdots<j_{M}\leq r$,
such that $p^{2}\sigma^{2}_{E_{i_{m}}} = q^{2}\sigma^{2}_{F_{j_{m}}}$, $m\in\{1,\dots,M\}$,
the relative proof is very similar to the situation in the simple case with $p^{2}\sigma^{2}_{E} = q^{2}\sigma^{2}_{F}$. The details here have been omitted.
$\hfill\blacksquare$

\textbf{\underline{Proof of Theorem 2}}.

Through Lemma 14, we have
\begin{eqnarray*}\label{T4.6}
\dfrac{(\hat{L}-L^{*})^{T}(\hat{L}-L^{*})}{q\hat{\sigma}^{2}}= O_{p}(p^{-1})\textbf{1}_{r\times r},~ \dfrac{\hat{L}^{T}(\hat{L}-L^{*})}{q\hat{\sigma}^{2}}=O_{p}(p^{-1})\textbf{1}_{r\times r},\\ \dfrac{(\hat{\Lambda}-\Lambda^{*})^{T}(\hat{\Lambda}-\Lambda^{*})}{p\hat{\sigma}^{2}} =O_{p}(p^{-1})\textbf{1}_{c\times c},~
\dfrac{ \hat{\Lambda}^{T}(\hat{\Lambda}-\Lambda^{*}) }{p\hat{\sigma}^{2}}=O_{p}(p^{-1})\textbf{1}_{c\times c}.
\end{eqnarray*}
Following the technique proof in Lemma 13, we obtain
\begin{eqnarray*}\label{T4.7}
\dfrac{\hat{\Lambda}^{T}X\hat{L}}{pq\hat{\sigma}^{4}} = O_{p}(p^{-1})\textbf{1}_{c\times r}.
\end{eqnarray*}
With expression of $\hat{\sigma}^{2} - \sigma^{*2}$ in Lemma 12, we have
\begin{eqnarray*}\label{T4.8}
&&\hat{\sigma}^{2} - \sigma^{*2}= -\dfrac{r\hat{\sigma}^{2}}{q} - \dfrac{c\hat{\sigma}^{2}}{p} + \dfrac{1}{pq}\sum_{i=1}^{p}\sum_{j=1}^{q}{(\epsilon_{ij}^{2}-\sigma^{*2})} + o_{p}(p^{-1}),\nonumber \\
&&\Rightarrow\left( 1+ \dfrac{r}{q} + \dfrac{c}{p} \right)\hat{\sigma}^{2} - \sigma^{*2}= \dfrac{1}{pq}\sum_{i=1}^{p}\sum_{j=1}^{q}{(\epsilon_{ij}^{2}-\sigma^{*2})} + o_{p}(p^{-1}).
\end{eqnarray*}
This leads to the asymptotic distribution of $\hat{\sigma}^{2}$,
\begin{eqnarray*}\label{T4.9}
\sqrt{pq} \left( \hat{\tilde{\sigma}}^2-\sigma^{*2} \right) &\stackrel{d}{\longrightarrow}& N \left( 0, 2\sigma^{*4} \right),\hat{\tilde{\sigma}}^{2} = \left( 1 +\dfrac{r}{q} +\dfrac{c}{p} \right)\hat{\sigma}^{2}.
\end{eqnarray*}
Recalling the estimating equation of $L$ in Subsection A.8,
the $m$th ($m\in \{1,\dots,q\}$) row of which can be written as
\begin{eqnarray}\label{T4.11}
\textbf{0}_{r}&=&\dfrac{1}{q\hat{\sigma}^{2}}\hat{L}_{m\cdot} + \dfrac{\hat{L}^{T}X^{T}X\hat{L}\hat{L}_{m\cdot} }{pq^{2}\hat{\sigma}^{6}} - \dfrac{\hat{\Psi}^{-1}_{F}\hat{L}^{T}X^{T}X\hat{L}\hat{L}_{m\cdot} }{pq^{3}\hat{\sigma}^{6}} - \dfrac{\hat{L}^{T}X^{T}X\hat{L}\hat{\Psi}^{-1}_{F}\hat{L}_{m\cdot} }{pq^{3}\hat{\sigma}^{6}}\nonumber \\
&&- \dfrac{\hat{L}^{T}X^{T}\hat{\Lambda}\hat{\Lambda}^{T}X\hat{L}\hat{L}_{m\cdot}}{p^{2}q^{2}\hat{\sigma}^{8}} - \dfrac{\left(X^{T}X\hat{L}\right)_{m\cdot} }{pq\hat{\sigma}^{4}} + \dfrac{\left(X^{T}X\hat{L}\hat{\Psi}^{-1}_{F}\right)_{m\cdot} }{pq^{2}\hat{\sigma}^{4}} + \dfrac{\left(X^{T}\hat{\Lambda}\hat{\Lambda}^{T}X\hat{L}\right)_{m\cdot}}{p^{2}q\hat{\sigma}^{6}} \nonumber \\
&&- \dfrac{\left(X^{T}\hat{\Lambda}\hat{\Lambda}^{T}X\hat{L}\hat{\Psi}^{-1}_{F}\right)_{m\cdot} }{p^{2}q^{2}\hat{\sigma}^{6}}  - \dfrac{\left(X^{T}\hat{\Lambda}\hat{\Psi}^{-1}_{E}\hat{\Lambda}^{T}X\hat{L}\right)_{m\cdot} }{p^{3}q\hat{\sigma}^{6}} + \left[\dfrac{X^{T}\hat{\Lambda}}{p\hat{\sigma}^{2}}\left(A^{(r,c)}_{\hat{\Psi}_{F},\hat{\Psi}_{E}} \circ \dfrac{\hat{\Lambda}^{T}X\hat{L}}{pq\hat{\sigma}^{4}}\right)\right]_{m\cdot}\nonumber \\
&&+ O_{p}(p^{-2})\hat{L}_{m\cdot} + O_{p}(p^{-2})\left(\hat{L}\textbf{1}_{r \times r}\right)_{m\cdot} + O_{p}(p^{-1})\left(\dfrac{\hat{L}\hat{L}^{T}X^{T}\hat{\Lambda}\hat{\Lambda}^{T}X\hat{L}}{p^{2}q^{2}\hat{\sigma}^{8}} \textbf{1}_{r\times r}\right)_{m\cdot}\nonumber \\
&&+ O_{p}(p^{-2})\dfrac{\left(X^{T}X\hat{L}\right)_{m\cdot} }{pq\hat{\sigma}^{4}} + O_{p}(p^{-2})\dfrac{\left(X^{T}\hat{\Lambda}\hat{\Lambda}^{T}X\hat{L}\right)_{m\cdot}}{p^{2}q\hat{\sigma}^{6}}\nonumber \\
&=& \dfrac{\hat{L}^{T}X^{T}X\hat{L}\hat{L}_{m\cdot} }{pq^{2}\hat{\sigma}^{6}} - \dfrac{\left(X^{T}X\hat{L}\right)_{m\cdot} }{pq\hat{\sigma}^{4}} + O_{p}(p^{-1})\textbf{1}_{r}.
\end{eqnarray}
With the expression of $\dfrac{\hat{L}^{T}X^{T}X\hat{L} }{pq^{2}\hat{\sigma}^{6}}$ in Lemma 9 and the above results, the final two terms of ($\ref{T4.11}$) can be simplified as
\begin{eqnarray}\label{T4.12}
&&\dfrac{\hat{L}^{T}X^{T}X\hat{L}\hat{L}_{m\cdot} }{pq^{2}\hat{\sigma}^{6}}= \dfrac{\hat{L}^{T}X^{T}X\hat{L} }{pq^{2}\hat{\sigma}^{6}}\cdot \hat{L}_{m\cdot}\nonumber  \\
&&= \left( \dfrac{ \hat{L}^{T}L^{*} }{q\hat{\sigma}^{2}}\dfrac{ F^{T}F }{p\hat{\sigma}^{2}}\dfrac{ L^{*T}\hat{L} }{q\hat{\sigma}^{2}} + \dfrac{\hat{L}^{T}L^{*}}{q\hat{\sigma}^{2}}\dfrac{F^{T}\Lambda^{*} }{p\hat{\sigma}^{2}} \dfrac{E^{T}\hat{L} }{q\hat{\sigma}^{2}} + \dfrac{\hat{L}^{T}L^{*}}{q\hat{\sigma}^{2}}\dfrac{F^{T}\epsilon \hat{L} }{pq\hat{\sigma}^{4}}+ \dfrac{\hat{L}^{T}E}{q\hat{\sigma}^{2}}\dfrac{\Lambda^{*T}F}{p\hat{\sigma}^{2}}\dfrac{L^{*T}\hat{L} }{q\hat{\sigma}^{2}}\right.\nonumber \\
&&\left. ~~ + \dfrac{\hat{L}^{T}E}{q\hat{\sigma}^{2}}\dfrac{\Lambda^{*T}\Lambda^{*} }{p\hat{\sigma}^{2}}\dfrac{E^{T}\hat{L} }{q\hat{\sigma}^{2}} + \dfrac{\hat{L}^{T}E}{q\hat{\sigma}^{2}}\dfrac{\Lambda^{*T}\epsilon \hat{L} }{pq\hat{\sigma}^{4}}+ \dfrac{\hat{L}^{T}\epsilon^{T}F}{pq\hat{\sigma}^{4}}\dfrac{L^{*T}\hat{L} }{q\hat{\sigma}^{6}} + \dfrac{\hat{L}^{T}\epsilon^{T}\Lambda^{*}}{pq\hat{\sigma}^{4}}\dfrac{E^{T}\hat{L} }{q\hat{\sigma}^{2}} + \dfrac{ \hat{L}^{T}\epsilon^{T}\epsilon \hat{L} }{pq^2\hat{\sigma}^{6}}\right) \cdot \hat{L}_{m\cdot}\nonumber \\
&&= \dfrac{F^{T}F}{p\hat{\sigma}^{2}}\hat{L}_{m\cdot} + O_{p}(p^{-1})\textbf{1}_{r},
\end{eqnarray}
and
\begin{eqnarray}\label{T4.13}
\dfrac{\left(X^{T}X\hat{L}\right)_{m\cdot} }{pq\hat{\sigma}^{4}}= \dfrac{F^{T}F}{p\hat{\sigma}^{2}} L^{*}_{m\cdot}+ \dfrac{F^{T}\Lambda^{*} }{p\hat{\sigma}^{2}}E_{m\cdot} + \dfrac{ \hat{L}^{T}E}{q\hat{\sigma}^{2}}E_{m\cdot}+  \dfrac{F^{T}\epsilon_{\cdot m}}{p\hat{\sigma}^{2}} + O_{p}(p^{-1})\textbf{1}_{r}.
\end{eqnarray}
Putting $(\ref{T4.12})$ and $(\ref{T4.13})$ into $(\ref{T4.11})$, we obtain
\begin{eqnarray*}\label{T4.14}
\textbf{0}_{r}&=&\dfrac{\hat{L}^{T}X^{T}X\hat{L}\hat{L}_{m\cdot} }{pq^{2}\hat{\sigma}^{6}} - \dfrac{\left(X^{T}X\hat{L}\right)_{m\cdot} }{pq\hat{\sigma}^{4}} + O_{p}(p^{-1})\textbf{1}_{r}\nonumber \\
&=&\dfrac{F^{T}F}{p\hat{\sigma}^{2}}\hat{L}_{m\cdot} - \dfrac{F^{T}F}{p\hat{\sigma}^{2}} L^{*}_{m\cdot} - \dfrac{F^{T}\Lambda^{*} }{p\hat{\sigma}^{2}}E_{m\cdot} - \dfrac{ \hat{L}^{T}E}{q\hat{\sigma}^{2}}E_{m\cdot} -  \dfrac{F^{T}\epsilon_{\cdot m}}{p\hat{\sigma}^{2}} + O_{p}(p^{-1})\textbf{1}_{r}\nonumber \\
&=&\dfrac{F^{T}F}{p\hat{\sigma}^{2}}\left(\hat{L}_{m\cdot} - L^{*}_{m\cdot}\right) - \dfrac{F^{T}\Lambda^{*} }{p\hat{\sigma}^{2}}E_{m\cdot} - \dfrac{ \hat{L}^{T}E}{q\hat{\sigma}^{2}}E_{m\cdot} -  \dfrac{F^{T}\epsilon_{\cdot m}}{p\hat{\sigma}^{2}} + O_{p}(p^{-1})\textbf{1}_{r},\nonumber \\
\Rightarrow&&\dfrac{F^{T}F}{p\hat{\sigma}^{2}}\left(\hat{L}_{m\cdot} - L^{*}_{m\cdot}\right)= \dfrac{F^{T}\epsilon_{\cdot m}}{p\hat{\sigma}^{2}} + \left(\dfrac{F^{T}\Lambda^{*} }{p\hat{\sigma}^{2}} + \dfrac{ \hat{L}^{T}E}{q\hat{\sigma}^{2}}\right)E_{m\cdot} +  O_{p}(p^{-1})\textbf{1}_{r},~~~~\nonumber \\
\Rightarrow&&\hat{L}_{m\cdot} - L^{*}_{m\cdot}= \Psi_{F}^{*-1}\dfrac{F^{T}\epsilon_{\cdot m}}{p} + \Psi^{*-1}_{F}\left(\dfrac{F^{T}\Lambda^{*} }{p} + \dfrac{ \hat{L}^{T}E}{q}\right)E_{m\cdot} +  O_{p}(p^{-1})\textbf{1}_{r}.~~~~
\end{eqnarray*}
From the conclusions of $\dfrac{ (\hat{L}-L^{*})^{T}E}{q\hat{\sigma}^{2}}$ in Lemma 13, we have
\begin{eqnarray*}\label{T4.15}
\sqrt{p} \left( \hat{L}_{m\cdot}-L_{m\cdot}^{*} \right) \stackrel{d}{\longrightarrow} N _{r}\left( 0, \Sigma_{L} \right),~m=1,\dots,q,
\end{eqnarray*}
where
\begin{eqnarray*}
\Sigma_{L} = \sigma^{*2}\Psi_{F}^{*-1} + \diag\left(\sum_{i=1}^{c}{\dfrac{\sigma^{*2}\sigma^{*2}_{E_{i}}(y\sigma^{*2}_{E_{i}} + \sigma^{*2}_{F_{1}})}{(\sigma^{*2}_{F_{1}} - \sigma^{*2}_{E_{i}})^2}} ,\dots, \sum_{i=1}^{c}\dfrac{\sigma^{*2}\sigma^{*2}_{E_{i}}(y\sigma^{*2}_{E_{i}} + \sigma^{*2}_{F_{r}})}{(\sigma^{*2}_{F_{r}} - \sigma^{*2}_{E_{i}})^2} \right).
\end{eqnarray*}
Similar results for  $\hat{\Lambda}_{k\cdot}$, $k=1,\dots,p$ can be obtained using the same techniques
\begin{eqnarray*}\label{T4.16}
&\sqrt{q} \left( \hat{\Lambda}_{k\cdot}-\Lambda_{k\cdot}^{*} \right) \stackrel{d}{\longrightarrow} N_{c}\left( 0, \Sigma_{\Lambda}\right),~k=1,\dots,p,
\end{eqnarray*}
where
\begin{eqnarray*}
\Sigma_{\Lambda} = \sigma^{*2}\Psi_{E}^{*-1} + \diag\left(\sum_{j=1}^{r}{\dfrac{\sigma^{*2}\sigma^{*2}_{F_{j}}(\sigma^{*2}_{F_{j}} + y\sigma^{*2}_{E_{1}})}{y(\sigma^{*2}_{E_{1}} - \sigma^{*2}_{F_{j}})^2}} ,\dots, \sum_{j=1}^{r}\dfrac{\sigma^{*2}\sigma^{*2}_{F_{j}}(\sigma^{*2}_{F_{j}} + y\sigma^{*2}_{E_{c}})}{y(\sigma^{*2}_{E_{c}} - \sigma^{*2}_{F_{j}})^2}  \right).
\end{eqnarray*}

For the asymptotic distributions of $\hat{\Psi}_{F}$ and $\hat{\Psi}_{E}$,
the estimating equation of $\sigma^{2}_{F_{j}}$, $j=1,\dots,r$, in Subsection A.8, is
\begin{eqnarray*}\label{T4.17}
&&\sum_{k=1}^{c}\dfrac{q\hat{\sigma}^{4}_{F_{j}}}{p(1+q\hat{\sigma}^{2}_{F_{j}} + p\hat{\sigma}^{2}_{E_{k}})} + \dfrac{(p-r)q\hat{\sigma}^{4}_{F_{j}}}{p(1+q\hat{\sigma}^{2}_{F_{j}})} - \dfrac{\hat{\sigma}^{4}_{F_{j}}\hat{L}^{T}_{j}X^{T}X\hat{L}_{j}}{p\hat{\sigma}^{4}(1+q\hat{\sigma}^{2}_{F_{j}})^{2}} \nonumber \\
&&+ \sum_{k=1}^{c}\dfrac{\hat{L}_{j}^{T}X^{T}\hat{\Lambda}_{k}\hat{\Lambda}_{k}^{T}X\hat{L}_{j}}{p^{2}\hat{\sigma}^{6}}\left[ \dfrac{\hat{\sigma}^{4}_{F_{j}}}{(1+q\hat{\sigma}^{2}_{F_{j}})^{2}} - \dfrac{\hat{\sigma}^{4}_{F_{j}}}{(1+q\hat{\sigma}^{2}_{F_{j}}+ p\hat{\sigma}^{2}_{E_{k}})^{2}} \right] = 0.
\end{eqnarray*}
With Lemma 10 and Lemma 11, we have
\begin{eqnarray*}\label{T4.18}
\hat{\sigma}^{2}_{F_{j}} - \dfrac{\hat{L}^{T}_{j}X^{T}X\hat{L}_{j}}{pq^{2}\hat{\sigma}^{4}} + O_{p}(p^{-1})=0,
\end{eqnarray*}
thus
\begin{eqnarray}\label{T4.19}
\diag\left(\hat{\Psi}_{F}\right) - \diag\left(\dfrac{\hat{L}^{T}X^{T}X\hat{L}}{pq^{2}\hat{\sigma}^{4}}\right) = O_{p}(p^{-1})\textbf{1}_{r}.
\end{eqnarray}
Moreover, since
\begin{eqnarray*}\label{T4.20}
\dfrac{\hat{L}^{T}X^{T}X\hat{L}}{pq^{2}\hat{\sigma}^{4}} = \dfrac{F^{T}F}{p} + O_{p}(p^{-1})\textbf{1}_{r\times r} = \Psi_{F}^{*} + \left(\dfrac{F^{T}F}{p} - \Psi_{F}^{*}\right)+ O_{p}(p^{-1})\textbf{1}_{r\times r},
\end{eqnarray*}
$(\ref{T4.19})$ then reduces to
\begin{eqnarray*}\label{T4.21}
\hat{\Psi}_{F} - \Psi_{F}^{*} = \dfrac{F^{T}F}{p} - \Psi_{F}^{*} + O_{p}(p^{-1})\textbf{1}_{r\times r},
\end{eqnarray*}
which implies that
\begin{eqnarray*}\label{T4.22}
\sqrt{p}\cdot \diag \left( \hat{\Psi}_{F} - \Psi_{F}^{*} \right) \stackrel{d}{\longrightarrow} N _{r}\left( 0, 2\Psi_{F}^{*2} \right).
\end{eqnarray*}
Similarly, we have
\begin{eqnarray*}\label{T4.23}
\sqrt{q} \cdot \diag \left( \hat{\Psi}_{E} - \Psi_{E}^{*} \right) \stackrel{d}{\longrightarrow} N _{c}\left( 0, 2\Psi_{E}^{*2} \right).
\end{eqnarray*}
$\hfill\blacksquare$

\scsection{References}

\begin{description}
\newcommand{\enquote}[1]{``#1''}
\expandafter\ifx\csname
natexlab\endcsname\relax\def\natexlab#1{#1}\fi

\bibitem[Adhikari(2007)]{Adhikari:2007}
Adhikari, S. (2007).
Matrix Variate Distributions for Probabilistic Structural Dynamics.
\textit{AIAA Journal}, 45, 1748-1762.

\bibitem[Allen and Tibshirani(2012)]{Allen:(2012)}Allen, G. I. and Tibshirani, R. (2012). Inference with Transposable Data: Modeling the Effects of Row and Column Correlations. \textit{Journal of the Royal Statistical Society: Series B}, 74(4), 721-743.

\bibitem[Anderson and Amemiya(1988)]{Anderson:(1988)}
Anderson, T. W. and  Amemiya, Y. (1988). The asymptotic normal distribution of estimators in factor analysis under general conditions. \textit{Annals of Statistics}, 16, 759-771.

\bibitem[Anderson and Rubin(1956)]{Anderson:(1956)}
Anderson, T. W. and Rubin, H. (1956). Statistical inference in factor analysis.
\textit{ Proceedings of the Third Berkeley Symposium on Mathematical Statistics
and Probability}, 5, 111-150.

\bibitem[Bai(2003)]{Bai:2003}
Bai, J. S. (2003). Inferential theory for factor models of large dimensions. \textit{Econometrica}, 71, 135-172.

\bibitem[Bai and Li(2012)]{Bai:2012}
Bai, J. S. and Li, K. (2012).
Statistical analysis of factor models of high dimension.
\textit{Annals of Statistics}, 40, 436-465.

\bibitem[Bai and Li (Sup)(2012)]{Bai(sup):2012}
Bai, J. S. and Li, K. (2012). Supplement to " Statistical analysis of factor models of high dimension." DOI:10.1214/11-AOS966SUPP.

\bibitem[Bickel and Levina(2008)]{Bickel:2008}
Bickel, P. and  Levina, E. (2008).
Regulatized estimation of large covariance matrices.
\textit{Annals of Statistics}. 36, 199-227.

\bibitem[Bolla(2001)]{Bolla:(2001)}
Bolla, M. (2001). Parallel factoring of strata. DOI:0.1109/ITI.2001.938028

\bibitem[Bolla~et~al(1998)]{Bolla:(1998)}
Bolla, M., Gy$\ddot{\text{o}}$rgy Michaletzky, G$\acute{\text{a}}$bor Tusn$\acute{\text{a}}$dy,  Ziermann, M. (1998). Extrema of sums of heterogeneous quadratic forms. \textit{Linear Algebra and Its Applications}, 269, 331-365.

\bibitem[Chen et al(2019)]{Chen:2019}
Chen, E.Y., Tsay, R. S., and Chen, R. (2019).
Constrained Factor Models for High-Dimensional Matrix-Variate Time Series.
\textit{Journal of the American Statistical Association},
DOI: 10.1080/01621459.2019.1584899

\bibitem[Dahl et al(2008)]{Dahl:2008}
Dahl, J., Vandenberghe, L., Roychowdhury, V. (2008).
Covariance selection for non-chordal graphs via
chordal embedding. \textit{Optim. Methods Softw}, 23, 501-520.

\bibitem[Ding and Cook(2018)]{Ding:2018}
Ding S. S., and Cook, R. D. (2018).
Matrix variate regressions and envelope models.
\textit{Journal of the Royal Statistical Society: Series B},  80(2), 387-408.

\bibitem[Doz et al(2012)]{Doz:(2012)}
Doz, C., Giannone, D. and Reichli, L. (2012). A Quasi Maximum Likelihood Approach for Large Approximate Dynamic Factor Models. \textit{The Review of Economics and Statistics} 94(4), 1014-1024.

\bibitem[Fan~et~al(2008)]{Fan:(2008)}
Fan, J., Fan, Y. and Lv, J. (2008).
High dimensional covariance matrix estimation using a factor model.
\textit{Journal of Econometrics}, 147, 186-197.

\bibitem[Fan et al(2011)]{Fan:(2011)}
Fan, J., Liao, Y., and Mincheva, M. (2011).
High Dimensional Covariance Matrix Estimation in Approximate Factor Models.
\textit{Annals of Statistics}, 39, 3320-3356.

\bibitem[Friedman et al(2007)]{Friedman:2007}
Friedman, J., Hastie, T. and Tibshirani, R. (2007)
Sparse inverse covariance estimation with the graphical Lasso.
\textit{Biostatistics}, 9, 432-441.

\bibitem[Gupta and Nagar(2000)]{Gupta:2000}
Gupta, A. and Nagar, D. (2000). Matrix Variate Distributions, Monographs
and Surveys in Pure and Applied Mathematics. Chapman Hall/CRC,
London.

\bibitem[Leng and Tang(2012)]{Leng:2012}
Leng, C. and Tang, C.Y. (2012). Sparse matrix graphical models.
\textit{Journal of the American Statistical Association}, 107 (499), 1187-1200.

\bibitem[Michael et al(2018)]{Michael:2018}
Michael Hornstein, Roger Fan, Kerby Shedden and  Shuheng Zhou. (2018).
Joint Mean and Covariance Estimation with Unreplicated Matrix-Variate Data.
\textit{Journal of the American Statistical Association}, 2018, DOI: 10.1080/01621459.2018.1429275.

\bibitem[Miller(1981)]{Miller:1981}
Miller, K. S. (1981). On the inverse of the sum of matrices.
\textit{Mathematics Magazine}, 54, 67-72.

\bibitem[Sheena and Gupta(2003)]{Sheena:2003}
Sheena, Y. and Gupta, A. (2003).
Estimation of the multivariate normal covariance matrix under some restrictions.
\textit{Statist. Decisions}, 21, 327-342.

\bibitem[Tsai et al(2016)]{Tsai:(2016)}
Tsai, H., Tsay, R., Lin, E. and Cheng, C. (2016). Doubly constrained factor models with applications. \textit{Statistica Sinica}, 26, 1453-1478.

\bibitem[Vandenberghe and Boyd(2004)]{Vandenberghe:2004}
Vandenberghe, L. and Boyd, S. (2004).
Convex Optimization. Cambridge University Press.

\bibitem[Wainwright et al(2006)]{Wainwright:2006}
Wainwright, M., Ravikumar, P. and Lafferty, J. D. (2006).
High-dimensional graphical model selection using $L_1$-regularized
logistic regression.
\textit{Proceedings of Advances in Neural information Processing Systems}.

\bibitem[Wang et al(2016)]{Wang Dong:2016}
Wang, D., Shen, H. and  Truong, Y. (2016).
Efficient dimension reduction for high-dimensional matrix-valued data.
\textit{Neurocomputing}, 190, 25-34.

\bibitem[Wang et al(2019)]{Wang:2019}
Wang, D., Liu, X. and Chen, R. (2019).
Factor models for matrix-valued high-dimensional time
series.
\textit{Journal of Econometric},  1, 231-248.

\bibitem[Werner et al(2008)]{Werner:2008}
Werner, K., Jansson, M. and  Stoica, P. (2008).
On estimation of covariance matrices with Kronecker product structure.
\textit{IEEE Trans. Signal Process}, 56, 478-491.

\bibitem[Won et al(2013)]{Won:2013}
Won, J., Lim, J., Kim, S., Rajaratnam, B. (2013).
Condition number regularized covariance estimation.
\textit{Journal of the Royal Statistical Society: Series B}, 75, 427-450.

\bibitem[Yuan(2009)]{Yuan:2009}
Yuan, M. (2009). Sparse inverse covariance matrix estimation via linear programming.
\textit{J. Mach. Learn. Res.}, 11, 2261-2286.

\bibitem[Yuan and Lin(2007)]{Yuan:2007}
Yuan, M. and Lin, Y. (2007).
Model selection and estimation in the Gaussian graphical model.
\textit{Biometrika}, 94, 19-35.

\bibitem[Zhou(2014)]{Zhou:2014}
Zhou, S. (2014).  Gemini: graph estimation with matrix variate normal instances.
\textit{Annals of Statistics}, 42, 532-562.

\end{description}

\end{singlespace}

\end{document}